# Cellular Connectivity for UAVs: Network Modeling, Performance Analysis and Design Guidelines

Mohammad Mahdi Azari[1], Fernando Rosas[2,3] and Sofie Pollin[1]
[1] Department of Electrical Engineering, KU Leuven, Belgium
[2] Centre of Complexity Science and Department of Mathematics, Imperial College London, UK
[3] Department of Electrical and Electronic Engineering, Imperial College London, UK

***Abstract*—The growing use of aerial user equipments (UEs) in various applications requires ubiquitous and reliable connectivity for safe control and data exchange between these devices and ground stations. Key questions that need to be addressed when planning the deployment of aerial UEs are whether the cellular network is a suitable candidate for enabling such connectivity, and how the inclusion of aerial UEs might impact the overall network efficiency. This paper provides an in-depth analysis of user and network level performance of a cellular network that serves both unmanned aerial vehicles (UAVs) and ground users in the downlink. Our results show that the favorable propagation conditions that UAVs enjoy due to their height often backfire on them, as the increased co-channel interference received from neighboring ground BSs is not compensated by the improved signal strength. When compared with a ground user in an urban area, our analysis shows that a UAV flying at 100 meters can experience a throughput decrease of a factor 10 and a coverage drop from 76% to 30%. Motivated by these findings, we develop UAV and network based solutions to enable an adequate integration of UAVs into cellular networks. In particular, we show that an optimal tilting of the UAV antenna can increase their coverage and throughput from 23% to 89% and from 3.5 b/s/Hz to 5.8 b/s/Hz, respectively, outperforming ground UEs. Furthermore, our findings reveal that depending on UAV altitude, the aerial user performance can scale with respect to the network density better than that of a ground user. Finally, our results show that network densification and the use of micro cells limit UAV performance. While UAV usage has the potential to increase area spectral efficiency (ASE) of cellular networks with moderate number of cells, they might hamper the development of future ultra dense networks.**

## I. Introduction

### A. Motivation

Drones or unmanned aerial vehicles (UAVs) are receiving great interest from both academia and industry. Thanks to recent developments in reliable and cost-effective hardware, the use of these platforms in diverse applications is rapidly becoming a reality, ranging from rescue missions to sensing application or even network deployment. As most of these applications depend on a reliable control of the UAVs and need a significant real-time data exchange, they require wireless technology that can guarantee an adequate connectivity between drones and ground-based networks [1], [2]. The key challenges that such technology needs to address are:

- *high coverage and continuous connectivity* to ensure continuous control and tracking of autonomous or piloted UAVs,
- *high throughput* to allow data exchange and video surveillance, and
- *low latency* to enable robust remote control or real-time applications such as event monitoring.

Additional features that are desirable from such technology include *secure communication* for data protection and privacy, *location verification* for traffic management, *licensed spectrum* for mission-critical applications, *system scalability* for supporting UAVs rapid growth, and *regulatory compliance* of UAV communication.

The wireless cellular network is a natural candidate for serving UAVs, as it is a mature technology which satisfies the above design objectives when serving ground users. In particular, using Long Term Evolution (LTE) technology with existing infrastructure offers several potential benefits such as flexible scheduling, resource management and multiple access mechanism [3]. This, in turn, could significantly reduce operational costs needed to enable the desired UAV connectivity. However, many parameters of the current instantiation of the cellular infrastructure and corresponding LTE technology have been optimized for ground users. For example, the fact that base station (BS) tilt their antennas down to the ground generates considerable antenna gain reduction for aerial users. Also, an spatial re-use factor designed for ground propagation conditions and inter-cell interference assumptions might not be applicable to UAVs. Moreover, recent studies have made clear the remarkable difference between ground-to-UAV communications and conventional ground-to-ground systems in terms of propagation conditions [4]–[6]. Therefore, there can exist a significant variation in the performance experienced by aerial users with respect to ground users when served by the ground cellular infrastructure.

In order to enable a successful use of cellular technology for serving UAVs, it is therefore critical to provide a clear characterization of the corresponding ground-to-UAV link properties and their relationship with principal network parameters. Moreover, it is crucial to explore potential coexistence conflicts that could exist between ground and aerial user equipments (UEs), and study the effect of various design factors on the quality of service (QoS) of both communities.

In summary, the key questions that drive this study are: Can current and future cellular networks be used to provide connectivity to aerial users? If yes, which are the major factors that could limit the performance for aerial users? Moreover, is it possible to leverage the flexibility of UAVs design to enhance the network performance? Finally, what are the main trade-offs when providing service to both aerial and ground users simultaneously?

### B. Related Works

The first theoretical investigations that consider UAV UEs (U-UEs) can be traced back to [7], [8] where the idea of *cellular-connected UAVs* are explored through the analysis of link coverage probability. By analyzing the impact of ground interfering BSs on aerial users, these works show that the network performance can be significantly improved by lowering BSs height and antenna tilt. To mitigate interference, furthermore, a learning based approach is proposed in [9].

In parallel to these analytical efforts, the feasibility of LTE-based UAV communication has also been examined via a set of field trials and simulations [10]–[12]. In [10] is shown by measurements that the number of detected cells tends to increase as the UAV height increases, due to more favorable fading conditions and less shadowing or obstruction of the electromagnetic propagation. These results are then used to feed numerical simulations to estimate the SINR for the downlink, while assuming free space propagation conditions and ignoring the effect of non-line-of-sight (NLoS) links and small-scale fading. Although insightful, it is not clear how to generalize these results to enable an efficient exploration of the impact and trade-offs of combinations of various system parameters. In [11], [12] the authors studied serving UAVs through cellular networks in the downlink for a suburban environment. Interestingly, several of findings reported in [7], [8] are corroborated by these measurements. Finally, a downlink service for aerial users through multi-user massive MIMO BSs is considered in [13], showing the performance enhancement as compared to existing single-user cellular network.

A complementary body of research considers UAVs not as cellular UEs but as aerial BSs or relays [4], [5], [14]–[21]. In [5], [14] the authors consider multiple aerial BSs serving ground users and optimize the network of UAVs for maximum performance. The results in [5] include an interference analysis of ground receivers harmed by distributed aerial transmitters at low altitudes, which is shown to be lower in denser environments for certain altitude ranges. To analyze UAV networks in an urban environment, a novel elevation angle and hence altitude dependent path loss model is proposed in [4] supported by the subsequent measurement based results [22]. In [21] an efficient 3D placement of drone cells are studied aiming to maximize the number of covered users. Optimal placement of UAV for maximum coverage and reliability is studied in [4], [17] by deriving outage probability for ground and aerial relaying respectively. A tutorial and general overview on UAV applications and corresponding challenges can be found in [15], [23], [24].

### C. Contributions

This paper presents a generic model to analyze the performance of UAVs served by a ground cellular network in terms of *coverage probability*, *achievable throughput* and *area spectral efficiency*. The proposed framework includes a dual-slope LoS/NLoS propagation model for both path loss and small scale fading, which are combined according to their probability of occurrence. The model also employs a generic distance and altitude dependent LoS probability that considers the effect of different types of urban environments. Therefore, following [4], this channel model reflects the fact that the path loss exponent and fading effect varies with link distance and altitude, while remaining unchanged within a certain range. The framework also includes the effect of the UAV antenna beamwidth and tilt angle, BSs height and various propagation environment and types of urban areas.

Using this framework, we derive exact expressions for the aforementioned performance metrics, and also approximations that allow insightful calculations in specific conditions. We analyze trends in user and network level performance for various combination of system parameters. The results are then used to evaluate the feasibility of ground and aerial users coexistence, and provide design guidelines for enabling an harmonious integration of UAVs in cellular networks.

A critical finding is that the combined effect of the signal loss due to BSs antenna downtilt and the increased LoS probability due to height introduces significant performance degradation to UAVs using omnidirectional antennas with respect to ground UEs (G-UEs). This fact, which is resulted from high amount of interference in the air, considerably reduces the operational range of UAV altitude. Motivated by this result, we explored the use of tilted directional antennas in UAV UEs which allows UAVs operate at higher altitudes. In particular, *we show that tilting the UAV antenna is notably beneficial for sparse to medium dense networks, being disadvantageous for very dense networks*. For instance, in one specific examined case corresponding to 4th generation (4G) specification, optimum UAV antenna tilting enhances the coverage probability from 23% to 89% and increases the achievable throughput from 3.5 b/s/Hz to 5.8 b/s/Hz. Interestingly, the optimum UAV antenna tilt is not the same for all performance metrics, which introduces a trade-off between coverage and throughput.

We further study optimal tier selection in a heterogeneous cellular network for UAV connectivity. *Optimal altitude-dependent tier selection* enhances link coverage from 10% to 80% and throughput from 0.25 b/s/Hz to 4 b/s/Hz at altitude of 100 m. We also address the optimal BSs deployment density from user and network perspective. For this, the overall effect of adding aerial users in the network is studied by considering scaling properties of user throughput and network area spectral efficiency. We found that when increasing network density a UAV UE can benefit considerably higher than a ground UE if appropriate choice of UAV altitude and antenna tilt angle is adopted. We recognized three regimes for aerial users when scaling the network with respect to its density, being different than that of a ground user. Finally, our findings show that network area spectral efficiency can be improved through the inclusion of aerial users.

### D. Paper Structure

The rest of paper is organized as follows. The network model is presented in Section II, including the system architecture, channel and blockages modeling in urban environments, and performance metrics considered for efficiency evaluation. Section III presents the main theoretical contributions, which include closed-form expressions for the performance metrics and approximations of reduced mathematical complexity. These results are then used in Section IV to study UAV connectivity in current and future of cellular networks. Finally, our main conclusions are summarized in Section V.

## II. Network Model

This section presents the considered cellular network and key modeling assumptions. In the sequel, Section II-A first presents general considerations about the network architecture. Then, Section II-B describes the channel model and Section II-C introduces blockages modeling in an urban area and the corresponding probability of LoS. Section II-D describes the BS association method and quantifies the corresponding signal-to-interference-plus-noise ratio (SINR) of each link, and finally Section II-E presents several performance metrics considered for user and network level efficiency evaluation.

### A. System Architecture

We consider a cellular network composed of ground base stations (BSs) and user equipments (UEs) representing both ground and aerial users. The ground BSs are randomly distributed according to a homogeneous Poisson point process (HPPP) $\Phi$ with density $\lambda\,\mathrm{BSs/Km^2}$, having $\mathrm{h_b}$ meters of height. Users are assumed to be located $\mathrm{h_u}$ meters above the ground, with $\mathrm{h_u} = 1.5\,\mathrm{m}$ for ground UEs. For a given BS, its ground distance to a specific UE's projection on the ground O is denoted by $r$, as illustrated in Figure 1.

The antenna radiation pattern of the BSs is assumed to be vertically directional and horizontally omnidirectional, implemented using multiple sector antennas. As these antennas are typically tilted down to help ground users coverage [3], the users above BS height are assumed to receive signals from the BSs sidelobe. The antenna gain of a BS is represented by $\mathrm{G_b}$, with $\mathrm{g_M}$ and $\mathrm{g_m}$ being the mainlobe and sidelobe gain.

We consider UAVs that are capable of tilting their directional antenna using a mechanical or electrical mechanism. The UAV's antenna is characterized by its opening angle $\phi_\mathrm{B}$ and tilt angle $\phi_\mathrm{t}$, as illustrated in Figure 1. The UAV antenna gain can be approximated by $\mathrm{G_u} = 29000/\phi_\mathrm{B}^2$ within the main lobe and zero outside of the main lobe [25]. Please note that, after tilting the antenna, a drone only receives signals coming from BSs within an elliptical section, denoted by $\mathcal{C}$ (c.f. Figure 1). When restricted to $\mathcal{C}$, the BSs also form an HPPP $\Phi_\mathcal{C}$ with the same density $\lambda$ [26]. The elliptical section can be described by its center $\mathrm{O_e} = (\mathrm{r_e}, 0)$, semi-major axis $\mathrm{r_M}$ and semi-minor axis $\mathrm{r_m}$, as illustrated in Figure 1.

Finally, the communication link length $d$ between a BS and a UE is $d = \sqrt{r^2 + \Delta_\mathrm{h}^2}$ where $\Delta_\mathrm{h} \triangleq \mathrm{h_u} - \mathrm{h_b}$. Please note that the tilted directional antenna for a UE is meaningful only if $\mathrm{h_u} > \mathrm{h_b}$. Accordingly, when $\mathrm{h_u} < \mathrm{h_b}$ it is assumed that the UE's antenna is omnidirectional.

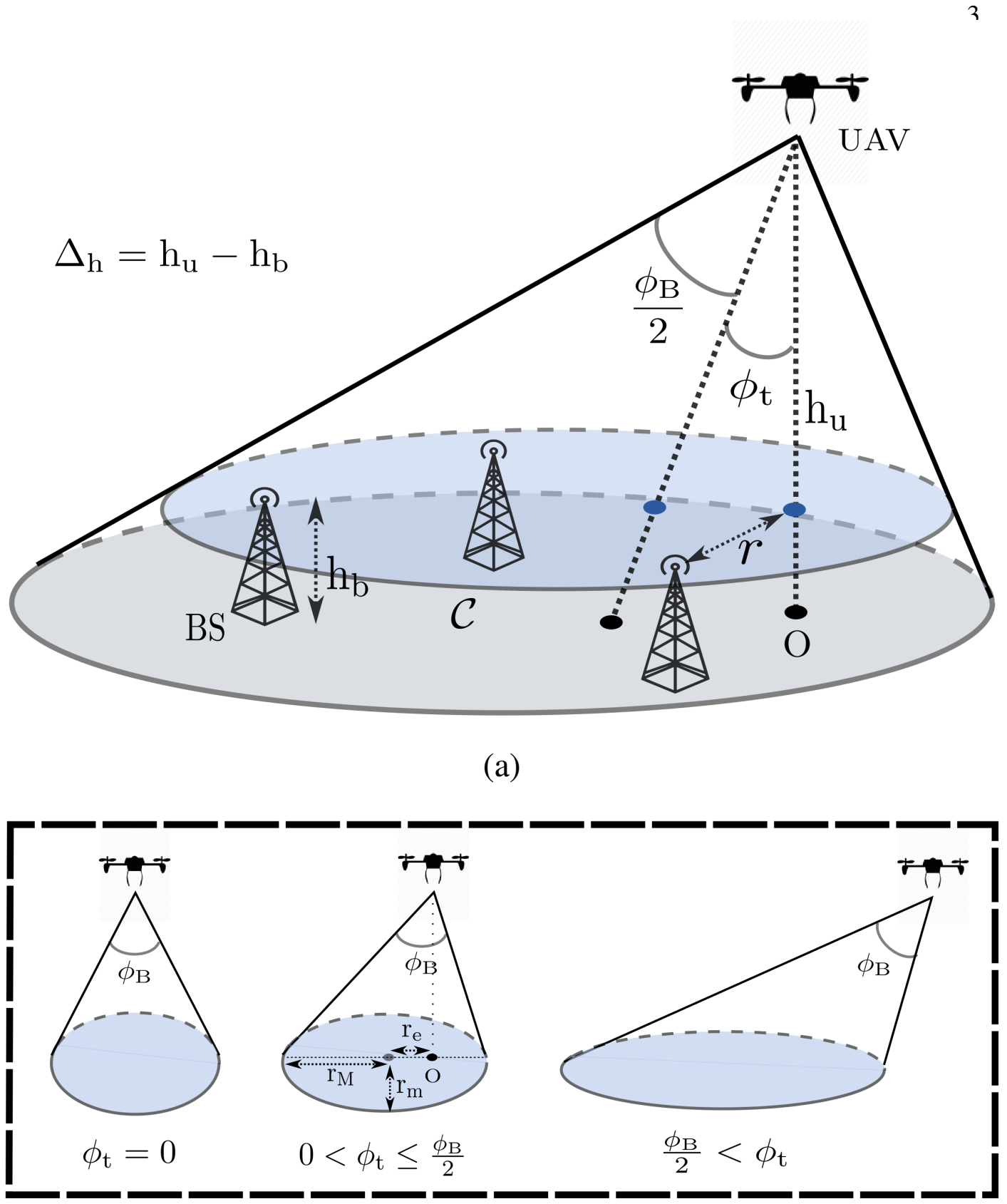

(a)

(b)

Fig. 1. Cellular connectivity for UAVs in 3D. (a) An illustration of the network geometry and parameters including UAV altitude $\mathrm{h_u}$, antenna tilt angle $\phi_\mathrm{t}$, antenna beamwidth $\phi_\mathrm{B}$, BS height $\mathrm{h_b}$, and the ground distance between a BS and the UAV $r$. (b) The UAV is equipped with a tilted directional antenna for performance enhancement. Because of this radiation pattern, the area seen by the UAV is modeled by an ellipse centered at a distance $\mathrm{r_e}$ from the projection of the UAV on the ground.

### B. Channel Model

In order to model the wireless link between a UE and a ground BS, the LoS and NLoS components are considered separately along with their probabilities of occurrence. The path loss $\zeta_\upsilon(r)$ for each LoS/NLoS components can be expressed as

$$\zeta_\upsilon(r) = \mathrm{A}_\upsilon d^{-\alpha_\upsilon} = \mathrm{A}_\upsilon \left(r^2 + \Delta_\mathrm{h}^2\right)^{-\alpha_\upsilon/2};\ \upsilon \in \{\mathrm{L}, \mathrm{N}\}, \quad (1)$$

where $\upsilon \in \{\mathrm{L}, \mathrm{N}\}$ indicates the condition of link being LoS ($\upsilon = \mathrm{L}$) or NLoS ($\upsilon = \mathrm{N}$). Moreover, $\alpha_\mathrm{L}$ and $\alpha_\mathrm{N}$ are the path loss exponents for the LoS and NLoS links respectively, and $\mathrm{A_L}$ and $\mathrm{A_N}$ are constants representing the path losses at the reference distance $d = 1$ for the LoS and NLoS cases respectively.

In our modeling wireless links undergo small scale fading with $\Omega_\upsilon$ being the fading power for the channel of condition $\upsilon$ which is either LoS or NLoS. Without loss of generality, we follow the convention $\mathbb{E}\{\Omega_\upsilon\} = 1$. For modeling the distribution of the channel gain $\Omega_\upsilon$ we use the Nakagami-m fading model, which can represent a wide range of fading environments [27]. Accordingly, $\Omega_\upsilon$ follows a Gamma distribution, whose

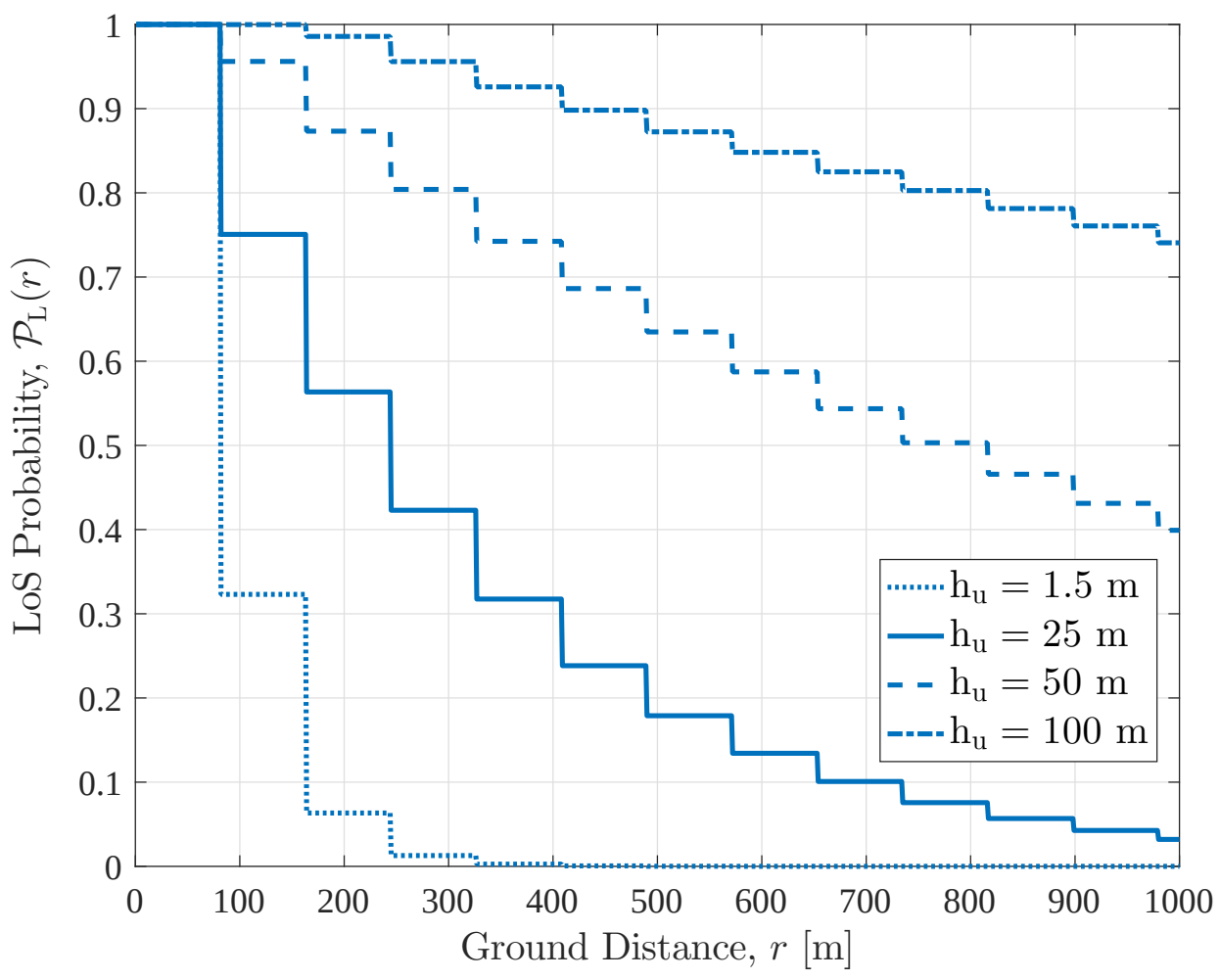

Fig. 2. LoS probability is an increasing function of $\rm h_u$ and a decreasing step function of r.

cumulative distribution function (CDF) can be expressed as [28]

$$F_{\Omega_v}(\omega) \triangleq \mathbb{P}[\Omega_v < \omega] = 1 - \sum_{k=0}^{m_v-1} \frac{(m_v\omega)^k}{k!} \exp(-m_v\omega), \quad (2)$$

where $m_v$ is the fading parameter assumed to be a positive integer for the sake of analytical tractability. Please note that, as a larger $m_v$ corresponds to a lighter fading, usually $m_{\rm L} > m_{\rm N}$ holds.

If a BS transmits with a power level $\rm P_{tx}$, the corresponding power received by a UE located at a ground distance $r$ is given by

$$P_{\rm rx}(r) = \rm P_{tx}\, G_{tot}\, \zeta_v(r)\, \Omega_v, \quad (3)$$

where $v$ is chosen depending if the link is LoS or NLoS, and $\rm G_{tot} = G_b\, G_u$ represents the total effect of transmitter and receiver antenna gains.

## C. *Blockages Modeling and LoS Probability*

We model an urban area as a set of buildings located in a square grid, following the method reported in [29]. The 3D blockages are characterized by three parameters: the fraction of total land area occupied by buildings, denoted by $a$, the mean number of buildings per $\rm km^2$, denoted by $b$, and the buildings height which are modeled as random variables that follow a Rayleigh distribution with an scale parameter $c$. Therefore, various environments (e.g. suburban, urban, dense urban or highrise urban) can be modeled by choosing an appropriate set of values $(a, b, c)$. Using this model, the proposed expression for the LoS probability between a given BS and UE can be expressed as (see Figure 2)

$$\mathcal{P}_{\rm L}(r) = \prod_{n=0}^{\rm M} \left[ 1 - \exp\left( - \frac{\left[ \rm h_b - \frac{(n+0.5)(h_b - h_u)}{m+1} \right]^2}{2c^2} \right) \right], \quad (4)$$

where $\rm M = \lfloor \frac{r\sqrt{ab}}{1000} - 1 \rfloor$. Correspondingly, the probability of NLoS is $\mathcal{P}_{\rm N}(r) = 1 - \mathcal{P}_{\rm L}(r)$.

For the sake of tractability we assume that the LoS probability of different communication links are independent, leaving the exploration of blockage correlations for future work. As a consequence of this assumption, one can decompose the PPP process $\Phi_{\mathcal{C}}$ in two independent but inhomogeneous PPP subprocesses: a PPP $\Phi_{\mathcal{C}}^{\rm L}$ of BSs that are in LoS condition with respect to the UE, henceforth called *LoS BSs*, and a PPP $\Phi_{\mathcal{C}}^{\rm N}$ of BSs in NLoS condition, henceforth called *NLoS BS*. The non-constant densities of $\Phi_{\mathcal{C}}^{\rm L}$ and $\Phi_{\mathcal{C}}^{\rm N}$ are denoted by $\lambda_{\rm L}(r) = \lambda\mathcal{P}_{\rm L}(r)$ and $\lambda_{\rm N}(r) = \lambda\mathcal{P}_{\rm N}(r)$, respectively. Therefore, $\Phi_{\mathcal{C}} = \Phi_{\mathcal{C}}^{\rm L} \cup \Phi_{\mathcal{C}}^{\rm N}$ and $\lambda = \lambda_{\rm L} + \lambda_{\rm N}$.

## D. *User Association Strategy and Link* SINR

We focus on the case where UEs establish communication links with the BS that provides the highest SINR. Interestingly, depending on the distribution of blockages, the serving BS can be either in LoS or NLoS condition, and it might not be the one closest to the UE.

Let us denote the ground distance between a given UE and the serving BS as $R_{\rm S}$. Then, using (3), the received power of the desired signal is given by $P_{\rm rx}(R_{\rm S})$, and the aggregate interference caused by all the other BSs within $\mathcal{C}$ can be written as

$$\begin{aligned} I &= \sum_{r \in \Phi_{\mathcal{C}} \setminus R_{\rm S}} P_{\rm rx}(r) \\ &= \sum_{r \in \Phi_{\mathcal{C}}^{\rm L} \setminus R_{\rm S}} \rm P_{tx}\, G_{tot}\, \zeta_L(r)\, \Omega_L + \sum_{r \in \Phi_{\mathcal{C}}^{\rm N} \setminus R_{\rm S}} \rm P_{tx}\, G_{tot}\, \zeta_N(r)\, \Omega_N. \end{aligned} \quad (5)$$

Above, the first and second term correspond to the total aggregate LoS and NLoS interference, respectively. Note that, due to the stochasticity of PPP, both $P_{\rm rx}(R_{\rm S})$ and $I$ are random variables.

Combining above results, the local instantaneous signal-to-interference-plus-noise ratio (SINR) for a serving BS can be finally stated as

$$\mathsf{SINR} = \frac{\rm P_{tx}\, G_{tot}\, \zeta_v(R_S)\, \Omega_v}{I + \rm N_0}; \quad v \in \{\rm L, N\}, \quad (6)$$

where $\rm N_0$ represents the noise power at the receiver's front-end.

## E. *Performance Metrics*

We evaluate the system performance from both user and network perspective using three metrics, which are described in the sequel. Please note that the distribution of the SINR, as defined in (6), is affected by many system parameters including the aerial UE altitude, $\rm h_u$. Therefore, all the metrics considered below depend on these parameters, although this might not be explicated in their functional definition.

*1) Coverage Probability:* denoted by $\mathcal{P}_{\rm cov}$, is defined by

$$\mathcal{P}_{\rm cov} \triangleq \mathbb{P}[\mathsf{SINR} > \rm T], \quad (7)$$

which can be written as $\mathcal{P}_{\rm cov} = \mathcal{P}_{\rm cov}(\rm h_u, T)$ as SINR depends on $\rm h_u$. The target value T is determined based on the user requirement and is related to the target rate $\rm R_t$ by $\rm T = 2^{R_t/BW} - 1$, where BW is the bandwidth allocated to each user. This performance metric reflects the *reliability* of the link between a UE and its associated BS in satisfying the target requirement. Additionally, this metric is useful to evaluate the reliability of a UAV link for *command and control (C&C)*, which includes critical information such as telemetry, identity, flight authorization, real time control for piloting, or navigation database and waypoint updates for autonomous UAVs. A reliable C&C link can enable UAV operations beyond visual LoS, being crucial for a safe UAV deployment and traffic management in future of UAV networks.

*2) Achievable Throughput (Channel Capacity):* denoted by $\mathcal{R}$, is the highest bit rate that a UE could obtain from the network. This metric is computed as

$$\mathcal{R} \triangleq \mathbb{E}\big[\log_2(1+\mathsf{SINR})\big] \quad \text{(b/s/Hz)}, \tag{8}$$

which depends on the user altitude $\mathrm{h_u}$ and the network density $\lambda$, and hence can be written as $\mathcal{R} = \mathcal{R}(\mathrm{h_u}, \lambda)$. The achievable throughput $\mathcal{R}$ quantifies performance in terms of average raw throughput, in contrast of the coverage probability that focuses on eventual events of service loss and other effects related to the quality of service and reliability of the link. As such, $\mathcal{R}$ and $\mathcal{P}_{\rm cov}$ provides complementary views over the BS to UE link quality.

*3) Area Spectral Efficiency (ASE):* denoted by $\mathcal{A}$, is a *network-level performance* metric that reflects the achievable data-rate per square meter. Let us denote as $\rho$ the ratio of UEs that correspond to aerial users. Then, on average $\lambda\rho$ BSs per square meter serve aerial users and $\lambda(1-\rho)$ serve ground users. Therefore, the average throughput per square meter can be calculated as

$$\mathcal{A} \triangleq \lambda[(1-\rho)\cdot\mathcal{R}(1.5,\lambda)+\rho\cdot\mathcal{R}(\mathrm{h_u},\lambda)] \quad \text{(b/s/Hz/Km}^2\text{)}, \tag{9}$$

where $\mathcal{R}(1.5,\lambda)$ and $\mathcal{R}(\mathrm{h_u},\lambda)$ corresponds to ground users at altitude of 1.5 m and aerial users at altitude of $\mathrm{h_u}$, respectively.

Through analyzing the ASE we aim to quantify the impact of adding aerial users to the network and investigate how the overall spectrum efficiency is affected when network resources are shared between ground and aerial users. In particular, we aim to study how the network scales in presence of aerial users and which consequences might be imposed by the inclusion of UAV UEs. Our analysis will enable to derive design guidelines for optimizing the ASE of networks with UAV and ground users.

## III. Performance Analysis

In this section we study the system described in the previous section with the performance metrics introduced in Section II-E. First, an exact expression for the coverage probability is derived in Subsection III-A. Then, in order to ease the numerical evaluations several approximations are proposed in Subsection III-B. Finally, the user achievable throughput and network ASE is calculated in Subsection III-C.

### A. Exact Coverage Probability

Given the system model and performance metrics defined earlier, Theorem 1 presents an exact expression for the coverage probability for a target rate $R_t$ corresponding to an $\mathsf{SINR}$ threshold T. In this theorem, we also obtain the distribution function of the serving BS distance $R_{\rm S}$, and characterize the aggregate interference through its Laplace transform.

**Theorem 1.** *The exact downlink coverage probability of a cellular-connected UAV equipped with a tilted directional antenna is obtained as*

$$\mathcal{P}_{\rm cov} = 2\sum_{\upsilon\in\{\mathrm{L,N}\}}\int_{\mathrm{r_0}}^{\mathrm{r_e+r_M}}\Big(\mathcal{P}^{\upsilon}_{\mathrm{cov}|R_{\rm S}}(\mathrm{r_S})\; f^{\upsilon}_{R_{\rm S}}(\mathrm{r_S}) \times \mathrm{r_S}[\pi+\varphi_1(\mathrm{r_S})-\varphi_2(\mathrm{r_S})]\Big)\,\mathrm{dr_S}, \tag{10}$$

*where* $\mathrm{r_0}$, $\mathrm{r_e}$, $\mathrm{r_M}$, $\varphi_1(r)$ *and* $\varphi_2(r)$ *are given in Table III in Appendix A. Above,* $f^{\upsilon}_{R_{\rm S}}(\mathrm{r_S})$ *is the probability density function (PDF) of the serving BS's distance* $R_{\rm S}$ *at an arbitrary angular coordinate within* $\mathcal{C}$*, which is calculated as*

$$f^{\upsilon}_{R_{\rm S}}(\mathrm{r_S}) = \lambda\mathcal{P}_{\upsilon}(\mathrm{r_S})\cdot e^{-2\lambda\left[\mathcal{I}^{\upsilon}_{\rm 1L}+\mathcal{I}^{\upsilon}_{\rm 1N}\right]};\;\; \upsilon\in\{\mathrm{L,N}\}, \tag{11}$$

*with*

$$\mathcal{I}^{\upsilon}_{1\xi} \triangleq \int_{\mathrm{r_0}}^{\mathrm{r}^{\upsilon}_{\xi}} r\mathcal{P}_{\xi}(r)[\pi+\varphi_1(r)-\varphi_2(r)]\;\mathrm{d}r;\;\; \xi\in\{\mathrm{L,N}\},$$

*and* $\mathrm{r}^{\upsilon}_{\xi}$ *can be found in Table III. Moreover,* $\mathcal{P}^{\upsilon}_{\mathrm{cov}|R_{\rm S}}$ *is the conditional coverage probability, given the serving BS of condition* $\upsilon$ *and its ground distance* $R_{\rm S}$*, which is derived as*

$$\mathcal{P}^{\upsilon}_{\mathrm{cov}|R_{\rm S}} = \sum_{k=0}^{m_\upsilon-1}(-1)^k q_k\cdot\frac{d^k}{dy^k_\upsilon}\mathcal{L}^{\upsilon}_{I|R_{\rm S}}(y_\upsilon);\;\; \upsilon\in\{\mathrm{L,N}\} \tag{12}$$

*where*

$$q_k \triangleq \frac{e^{-\mathrm{N_0}y_\upsilon}}{k!}\sum_{j=k}^{m_\upsilon-1}\frac{\mathrm{N}_0^{j-k}y^j_\upsilon}{(j-k)!}, \tag{13}$$

$$y_\upsilon \triangleq \frac{m_\upsilon\mathrm{T}}{\mathrm{P_{tx}}\,\mathrm{G_{tot}}\,\zeta_\upsilon(\mathrm{r_S})}, \tag{14}$$

*and* $\mathcal{L}^{\upsilon}_{I|R_{\rm S}}(y_\upsilon)$ *is the Laplace transform of the conditional aggregate interference* $I|R_{\rm S}$ *evaluated at* $y_\upsilon$ *for the serving BS of condition* $\upsilon$*. Finally,* $\mathcal{L}^{\upsilon}_{I|R_{\rm S}}(\cdot)$ *can be expressed as*

$$\mathcal{L}^{\upsilon}_{I|R_{\rm S}}(y_\upsilon) = e^{-2\lambda[\mathcal{I}^{\upsilon}_{\rm 2L}+\mathcal{I}^{\upsilon}_{\rm 2N}]}, \tag{15}$$

*where*

$$\mathcal{I}^{\upsilon}_{2\xi} \triangleq \int_{\mathrm{r}^{\upsilon}_{\xi}}^{\mathrm{r_e+r_M}} r\mathcal{P}_{\xi}(r)\,[1-\Upsilon_{\xi}(r,y_\upsilon)] \times[\pi+\varphi_1(r)-\varphi_2(r)]\,\mathrm{d}r \tag{16}$$

*with* $\mathrm{r}^{\upsilon}_{\xi}$ *given in Table III and*

$$\Upsilon_{\xi}(r,y_\upsilon) \triangleq \left(\frac{m_\xi}{m_\xi+y_\upsilon\mathrm{P_{tx}}\,\mathrm{G_{tot}}\,\zeta_\xi(r)}\right)^{m_\xi}. \tag{17}$$

*Proof.* See Appendix A. □

Note that Theorem 1 can also be used to find the coverage probability of a user equipped with an omnidirectional antenna. For doing this, one needs to use above equations replacing $\mathrm{r_e} = \mathrm{r_0} = 0$, $\mathrm{r_M} = \infty$, and $\varphi_1(r) = \varphi_2(r)$.

### B. Approximations for UAV Coverage Probability

In this subsection we explore two methods for finding approximations that reduce the complexity of the equations involved in Theorem 1. These methods allow to focus into factors that play major roles into the achievable performance of UAV UEs. To this end, in the sequel we first show when NLoS links and noise effects can be neglected, and then use this result to build simplified approximations for the coverage probability.

*1) Discarding NLoS and noise effects:* For a UAV UE flying above the BS height, the number of LoS BSs is considerably higher than that of a ground UE. These BSs are likely to be closer to the UAV since the probability of LoS increases as the distance decreases. Therefore, due to lower path loss exponent and shorter distances, the received power of those nearby LoS BSs will be dominant as compared to the aggregate received power from NLoS BSs. Furthermore, the noise effect could also be neglected due to the high amount of aggregate interference power received from the LoS BSs. These facts motivate us to consider the case that the communication link with UAV is LoS and interference limited in order to simplify expressions as follows.

**Proposition 1.** *The UAV-UE coverage performance $\mathcal{P}_{\rm cov}$ can be significantly simplified by eliminating NLoS links and thermal noise effects as*

$$\mathcal{P}_{\rm cov} \approx 2\int_0^{\rm r_e+r_M} \mathcal{P}^{\rm L}_{{\rm cov}|R_{\rm S}}({\rm r_S})\ f^{\rm L}_{\rm R_S}({\rm r_S})[\pi+\varphi_1({\rm r_S})-\varphi_2({\rm r_S})]\ {\rm r_S}{\rm dr_S},$$

*where*

$$f^{\rm L}_{\rm R_S}({\rm r_S}) \approx \lambda\mathcal{P}_{\rm L}({\rm r_S})\cdot e^{-2\lambda\mathcal{I}^{\rm L}_{\rm 1L}}, \tag{18}$$

$$\mathcal{P}^{\rm L}_{{\rm cov}|R_{\rm S}} \approx \sum_{k=0}^{m_{\rm L}-1}\frac{(-y_{\rm L})^k}{k!}\cdot\frac{d^k}{dy_{\rm L}^k}\mathcal{L}^{\rm L}_{I|R_{\rm S}}(y_{\rm L}), \tag{19}$$

$$\mathcal{L}^{\rm L}_{I|R_{\rm S}}(y_{\rm L}) \approx e^{-2\lambda\mathcal{I}^{\rm L}_{\rm 2L}}. \tag{20}$$

*Proof.* The above is a direct consequence of using $\mathcal{P}_{\rm N}\approx 0$ and $N_0\approx 0$ in Theorem 1. □

Proposition 1 provides an elegant expression to estimate $\mathcal{P}_{\rm cov}$, which is obtained by neglecting several terms that correspond to NLoS links, including $\mathcal{I}^{\rm N}_{1\xi}$ and $\mathcal{I}^{\rm L}_{\rm 1N}$. Moreover, the convolved calculation of $\mathcal{L}^{v}_{I|R_{\rm S}}$ in Theorem 1 is significantly simplified, as $\mathcal{I}^{\rm L}_{\rm 2N}$ is removed from the calculation of the Laplace transform. The accuracy of this approximation is explored in Section IV.

*2) Moment Matching:* To enable a more simplified calculation of the coverage probability using Proposition 1, in the sequel, we pursue a parametric estimation of the interference statistics based on a Gamma distribution. This is done by a moments matching method, which allows us to find a closed-form approximation of the interference Laplace transform. Using this approximation, the integral and derivatives corresponding to the interference Laplace transform are eliminated, greatly simplifying the calculations.

As a first step, we provide expressions for the mean and variance of the interference.

**Lemma 1.** *First and second moments of the interference PDF can be calculated as*

$$\mu_{I|R_{\rm S}} = 2\lambda{\rm P_{tx}G_{tot}}\int_{\rm r_S}^{\rm r_e+r_M} r\mathcal{P}_{\rm L}(r)\zeta_{\rm L}(r)[\pi+\varphi_1(r)-\varphi_2(r)]\,{\rm d}r \tag{21}$$

*and*

$$\sigma^2_{I|R_{\rm S}} = 2\lambda({\rm P_{tx}G_{tot}})^2\left(\frac{m_{\rm L}+1}{m_{\rm L}}\right) \times\int_{\rm r_S}^{\rm r_e+r_M} r\mathcal{P}_{\rm L}(r)\zeta^2_{\rm L}(r)[\pi+\varphi_1(r)-\varphi_2(r)]\,{\rm d}r. \tag{22}$$

*Proof.* See Appendix B. □

Note that $\mathcal{P}_{\rm L}(r)$, as defined in (4) in Section II-C, is a decreasing step function of $r$ as illustrated in Figure 2. In particular, for an arbitrary non-negative integer value of k and any $r\in[{\rm r_k},{\rm r_{k+1}}]$ with ${\rm r_k}=1000({\rm k}+1)/\sqrt{ab}$, we have $\mathcal{P}_{\rm L}(r)={\rm p_k}$ where ${\rm p_k}$ is a fixed value obtained from (4) by replacing ${\rm M}={\rm k}$. Therefore, the integrals in Lemma 1 can be obtained through the following corollary for $\phi_{\rm t}=0$. This, in turn, corresponds to the case where the UAV antenna is pointing directly down, or when the antenna is omnidirectional.

**Corollary 1.** *For the case that $\phi_{\rm t}=0$, the first and second moments of interference can be obtained as*

$$\mu_{I|R_{\rm S}} = \frac{\pi\lambda{\rm P_{tx}G_{tot}A_L}}{0.5\,\alpha_{\rm L}-1} \times\sum_{k={\rm i}}^{\rm j}{\rm p_k}\left[({\rm r_k^2}+\Delta_{\rm h}^2)^{1-0.5\,\alpha_{\rm L}}-({\rm r_{k+1}^2}+\Delta_{\rm h}^2)^{1-0.5\,\alpha_{\rm L}}\right], \tag{23}$$

*and*

$$\sigma^2_{I|R_{\rm S}} = \frac{\lambda({\rm P_{tx}G_{tot}A_L})^2}{\alpha_{\rm L}-1}\left(\frac{m_{\rm L}+1}{m_{\rm L}}\right) \times\sum_{k={\rm i}}^{\rm j}{\rm p_k}\left[({\rm r_k^2}+\Delta_{\rm h}^2)^{1-\alpha_{\rm L}}-({\rm r_{k+1}^2}+\Delta_{\rm h}^2)^{1-\alpha_{\rm L}}\right], \tag{24}$$

*in which* ${\rm i}=\lfloor\frac{{\rm r_S}\sqrt{ab}}{1000}-1\rfloor$, ${\rm j}=\lfloor\frac{{\rm r_M}\sqrt{ab}}{1000}-1\rfloor$, ${\rm r_k}=1000({\rm k}+1)/\sqrt{ab}$ *except that* ${\rm r_i}={\rm r_S}$ *and* ${\rm r_{j+1}}={\rm r_M}$, *and* ${\rm p_k}$ *is a fixed value obtained from* (4) *by replacing* ${\rm M}={\rm k}$.

*Proof.* Please find Appendix C. □

A Gamma approximation of $I$ can be characterized by its scale and shape parameters, denoted as $\beta_2,\beta_1$ respectively. The relationship between these parameters and the interference's mean and variance is [28]

$$\beta_1=\frac{\sigma^2_{I|R_{\rm S}}}{\mu_{I|R_{\rm S}}},\ \beta_2=\frac{\mu^2_{I|R_{\rm S}}}{\sigma^2_{I|R_{\rm S}}}. \tag{25}$$

Moreover, the Laplace transform of Gamma distribution can be written as [28]

$$\mathcal{L}^{\rm L}_{I|R_{\rm S}}(y_{\rm L})=(1+\beta_1 y_{\rm L})^{-\beta_2}. \tag{26}$$

Using this approximation, one can directly obtain the coverage probability of UAV communication link using the next proposition.

**Proposition 2.** *The statistics $I$ can be approximated by a Gamma distribution with scale and shape parameters $\beta_2$ and $\beta_1$, and hence the conditional coverage performance of a UAV-UE can be approximated by*

$$\mathcal{P}^{\rm L}_{{\rm cov}|R_{\rm S}} \approx \frac{1}{\Gamma(\beta_2)}\sum_{k=0}^{m_{\rm L}-1}\frac{(\beta_1 y_{\rm L})^k}{k!}\,\Gamma(\beta_2+k)\,(1+\beta_1 y_{\rm L})^{-\beta_2-k}, \tag{27}$$

*where $\Gamma(\cdot)$ is the complete gamma function*[1].

*Proof.* The result is obtained by substituting (26) in (19). □

[1]Please note that using the property of gamma function we have $\Gamma(\beta_2+k)/\Gamma(\beta_2)=\prod_{i=0}^{k-1}(\beta_2+i)$.

### C. Achievable Throughput and ASE Analysis

The achievable throughput of a typical UE introduced in Section II-E can be obtained as

$$\mathcal{R} \triangleq \mathbb{E}\left[\log_2(1+\mathsf{SINR})\right] \tag{28}$$
$$= \frac{1}{\ln 2}\int_0^\infty \frac{\mathcal{P}_{\rm cov}({\rm h_u}, t)}{1+t}\,{\rm d}t$$
$$\approx \frac{1}{\ln 2}\sum_{n=1}^{\rm K} \frac{\mathcal{P}_{\rm cov}({\rm h_u}, t_n)}{1+t_n}\cdot \frac{\pi^2 \sin\left(\frac{2n-1}{2{\rm K}}\pi\right)}{4{\rm K}\cos^2\left[\frac{\pi}{4}\cos\left(\frac{2n-1}{2{\rm K}}\pi\right)+\frac{\pi}{4}\right]}. \tag{29}$$

Above, the last equation is an approximation to facilitate numerical calculations and follows the Gauss-Chebyshev Quadrature (GCQ) rule, whose free parameter $K$ is to be chosen large enough for a high accuracy of approximation [30]. Also, $t_n$ is a shorthand notation for

$$t_n = \tan\left[\frac{\pi}{4}\cos\left(\frac{2n-1}{2{\rm K}}\pi\right)+\frac{\pi}{4}\right]. \tag{30}$$

Finally, ASE is obtained by a direct substitution of $\mathcal{R}({\rm h_u}, \lambda)$ from (29) into (9).

## IV. System Design: Study Cases and Discussion

The goal of this section is to use the theoretical tools developed in the Section III to enable a qualitative and quantitative understanding of key design parameters for the inclusion of aerial users in cellular networks. For this aim, we ran network simulations and numerical evaluations using the parameter values listed in Table I.

In the sequel, Section IV-A validates the analytical expressions obtained for coverage probability in Theorem 1, and also explores the accuracy of the approximations proposed in Propositions 1 and 2. Then, Subsection IV-B studies the impact of different system parameters on the coverage probability and expected throughput of the UAV communication link, and compares the results with ground users performance. Finally, we study UAV connectivity performance by considering heterogeneous networks in Subsection IV-C and network densification in Subsection IV-D.

TABLE I. Default values for simulation and numerical evaluation.

| Parameter | Value |
|---|---|
| $(\alpha_{\rm L}, \alpha_{\rm N})$ | $(2.09\,,\,3.75)$ |
| $({\rm A_L}, {\rm A_N})$ | $(-41.1\,,\,-32.9)$ dB |
| $(m_{\rm L}, m_{\rm N})$ | $(3\,,\,1)$ |
| ${\rm P_{tx}}$ | 46 dBm |
| $(a, b, c)$ | $(0.3\,,\,500\,,\,15)$ @Urban |
| $\lambda$ | 10 BSs/Km$^2$ |
| $({\rm g_M}, {\rm g_m})$ | $(10\,,\,0.5)$ |
| ${\rm h_u}$ | 100 m |
| ${\rm h_b}$ | 25 m |
| BW | 200 KHz |
| ${\rm R_t}$ | 100 Kbps |
| T | $2^{\rm R_t/BW}-1=-3.8$ dB |

### A. Analysis of Accuracy

Figure 3 shows the complementary cumulative distribution function (CCDF) of the received SINR at different altitudes for users equipped with omnidirectional antennas. Simulation results are obtained by averaging over $10^5$ network realizations. Simulation results are in perfect correspondence with the expressions presented in Theorem 1, and further illustrate the accuracy of approximations presented in Propositions 1 and 2. As noted in Section III-B, the accuracy of the approximations relies on the relatively higher probability of LoS with BSs detected by a UAV, which reduces the impact of NLoS links and noise. Please note that for ground users this approximation is not valid and, hence, the corresponding curve is not included in the figure.

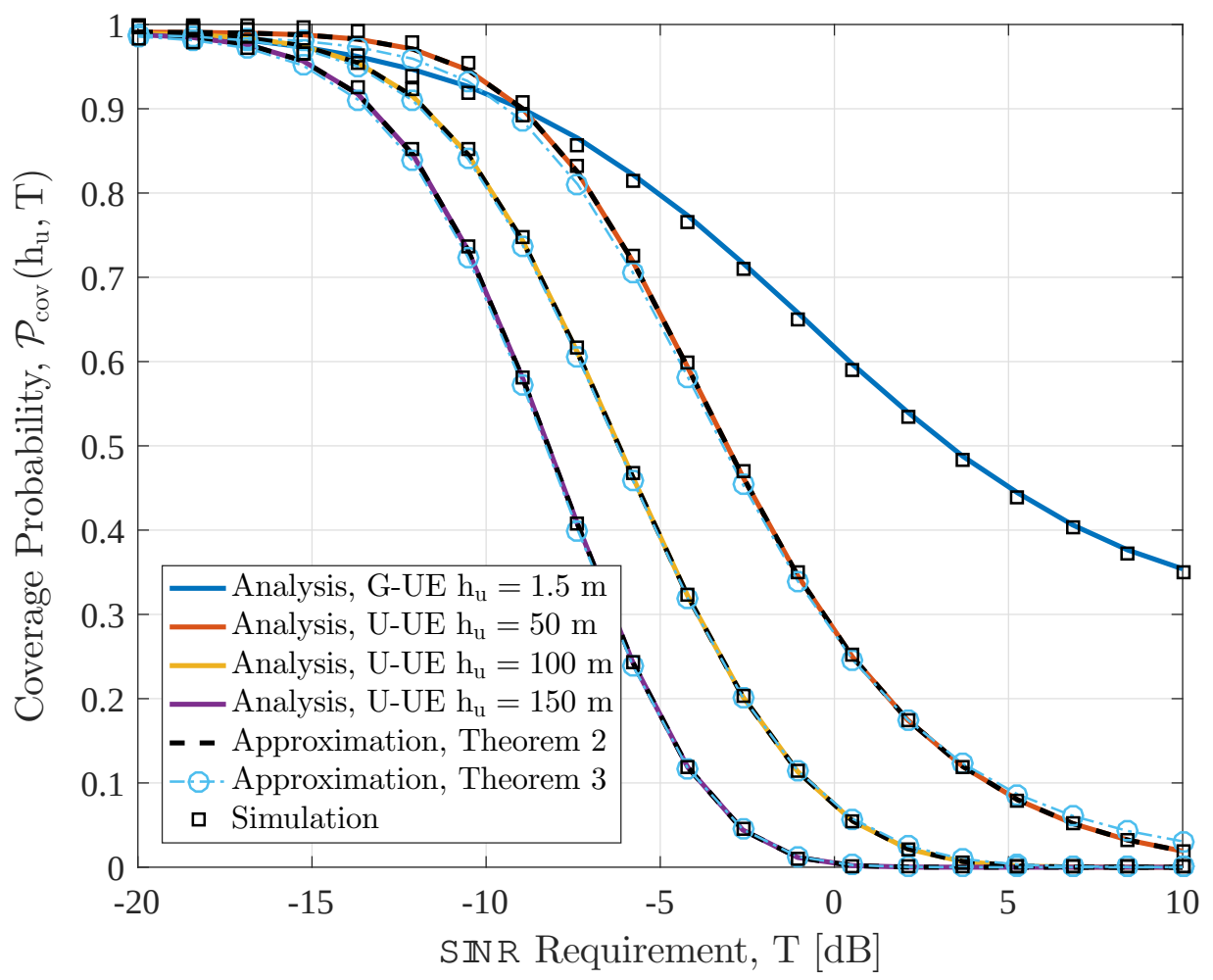

Fig. 3. Results from analytical expressions in Theorem 1 and network simulation are in a good conformity. Moreover, proposed approximations in Propositions 1 and 2 for UAV UEs (U-UEs) precisely follow the observed trend in exact coverage probability. However, these approximations are not included for ground UEs (G-UEs) since NLoS links play prominent role for G-UEs and hence they are away from the area where the approximations are supposed to be accurate. In general, the coverage of UAV communication link at higher altitude drops significantly when it is equipped with omnidirectional antenna.

Indirectly, this figure also reveals the effect of altitude on the SINR distribution. In effect, as the UAV altitude increases the distribution of SINR becomes more concentrated. This effect is due to the fact that the total signal power received from LoS links grows and dominates over the multipath scatterers. In contrast, it can also be seen how multipath scatterers play a prominent role in determining the SINR distribution for a G-UE due to the low probability of LoS with BSs.

### B. Design Parameters

This subsection provides an in-depth analysis of the impact of various design parameters on the quality of the UAV communication link with BSs.

*1) **Impact of Altitude**:* Our results show in general that *there is an optimum altitude where the performance of cellular-connected UAV is maximized*, as illustrated in Figure 4. In fact, at the optimum altitude the LoS probability of the serving BS is enhanced compared to the ground level while keeping most of the interfering BSs in NLoS condition. However, at higher altitudes, the interfering BSs start to become more LoS and hence significantly reduce the communication link quality. Our results show that even when considering sidelobe attenuation for the radiated signals, the impact of interference at relatively high altitudes is significant, introducing considerable degradation in the performance of UAV communication link. The strong detrimental effect of the interference makes the optimal UAV altitude $\mathrm{h_u}$ to be relatively low, as illustrated in Figure 4.

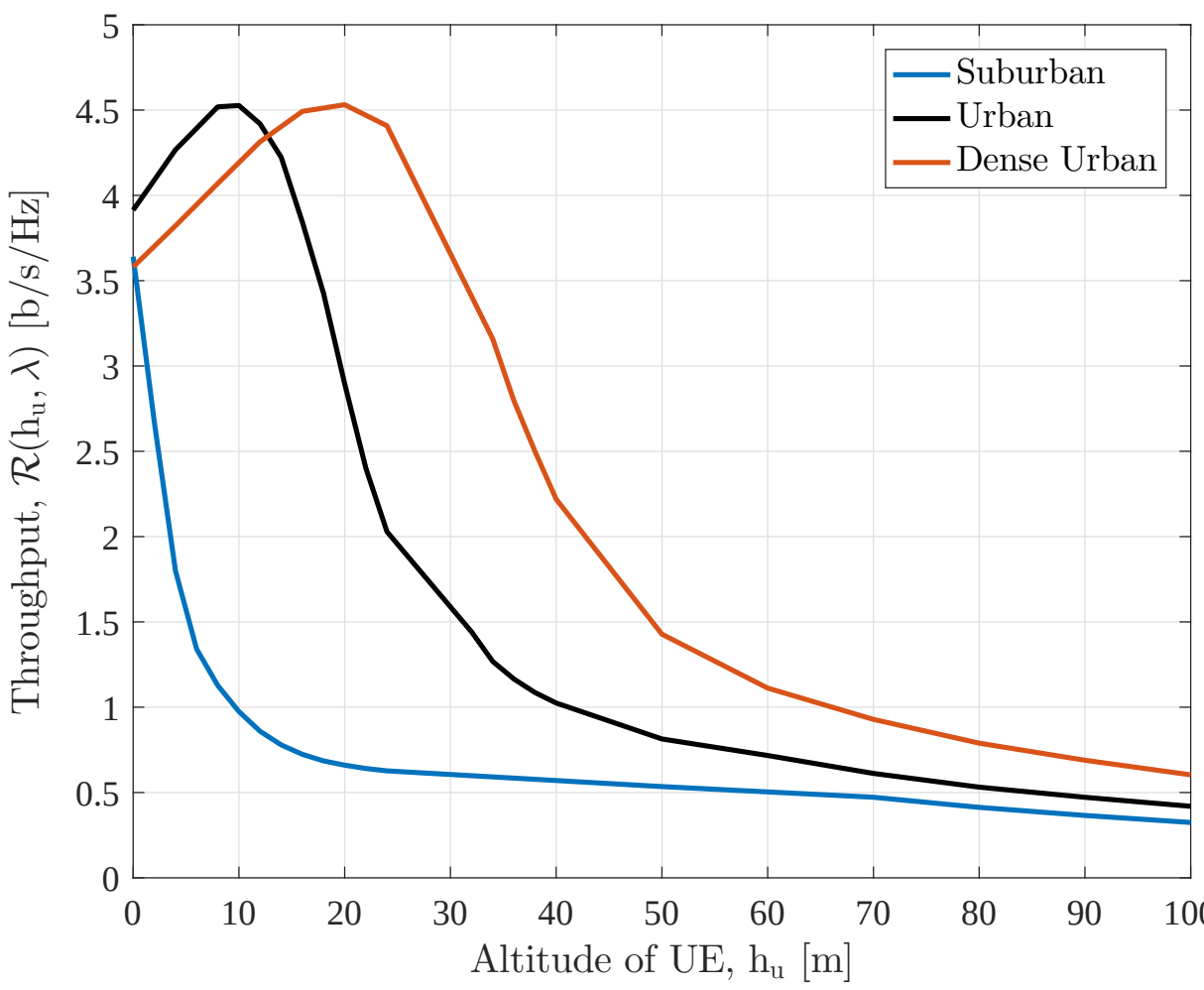

Fig. 4. In an urban area achievable throughput (channel capacity) of an omnidirectional UAV at altitude of 100 m is only 12% of a ground user and 10% of its maximum at the optimum altitude, *i.e.* $\mathrm{h_u}$ =10 m. In this case feasible operational range of UAV altitude is very limited.

For example, in an urban environment throughput is maximized at $\mathrm{h_u} = 10\,\mathrm{m}$, and at a higher altitude of $\mathrm{h_u} = 100\,\mathrm{m}$ the UAV achieves only 10% of its maximum throughput. These results are consistent with the findings reported for coverage probability in [8]. In dense urban scenarios experiencing more blockages, a higher optimal UAV altitude is found. The situation is however different in suburban scenarios where there are not so many blockages, and UAVs should fly as low as possible (see Figure 4).

*2) **Impact of the Environment:*** When comparing different urban areas, Table II shows that the impact of altitude on the coverage probability is more severe in less obstructed areas. For instance, the link coverage decreases from 90% at ground level to only 4% at 150 m in a suburban area, which is more drastic than the corresponding drop from 76% to 10% in an urban area. Interestingly, a ground user has the best coverage probability in suburban scenarios, while a UAV user at 50 m or higher is best served in urban scenarios. *Network deployment strategies are hence expected to be different for ground and aerial users.*

TABLE II. Link coverage of a cellular-connected UAV with an omnidirectional antenna in two types of urban areas. A target rate of $\mathrm{R_t} = 100$ Kbps is assumed for coverage assessment.

| UAV Altitude $\mathrm{h_u}$ [m] | Assigned BW [KHz] | Coverage @Suburban | Coverage @Urban |
|---|---|---|---|
| 1.5 | 200 | 90% | 76% |
| 50 | 200 | 34% | 54% |
| 100 | 200 | 20% | 30% |
| 150 | 200 | 4% | 10% |
| 1.5 | 400 | 97% | 85% |
| 50 | 400 | 60% | 82% |
| 100 | 400 | 48% | 60% |
| 150 | 400 | 28% | 39% |

It is direct to see that, as the UAV link is interference limited, the achievable throughput $\mathcal{R}$ grows linearly with the assigned bandwidth (BW). In contrast, the link coverage probability is affected by the BW in a non-trivial way. Table II illustrates this dependency through four sample altitudes in two different environments, focusing on a fixed target rate of $\mathrm{R_t} = 100\,\mathrm{Kbps}$ for UAV C&C [31]. Results show that UAVs with omnidirectional antenna endure considerably lower link coverage than a ground user particularly in less obstructed areas. Doubling the bandwidth for a given target rate reduces the SINR threshold constraint and hence significantly improves the coverage probability, e.g., the coverage probability is improved almost 4 times for a UAV at 150 m in an urban setting. On the other hand, the same BW doubling enables a growth of only %8 for UE at ground level. Therefore, in general, our results suggest that an increase in bandwidth is more efficient for aerial

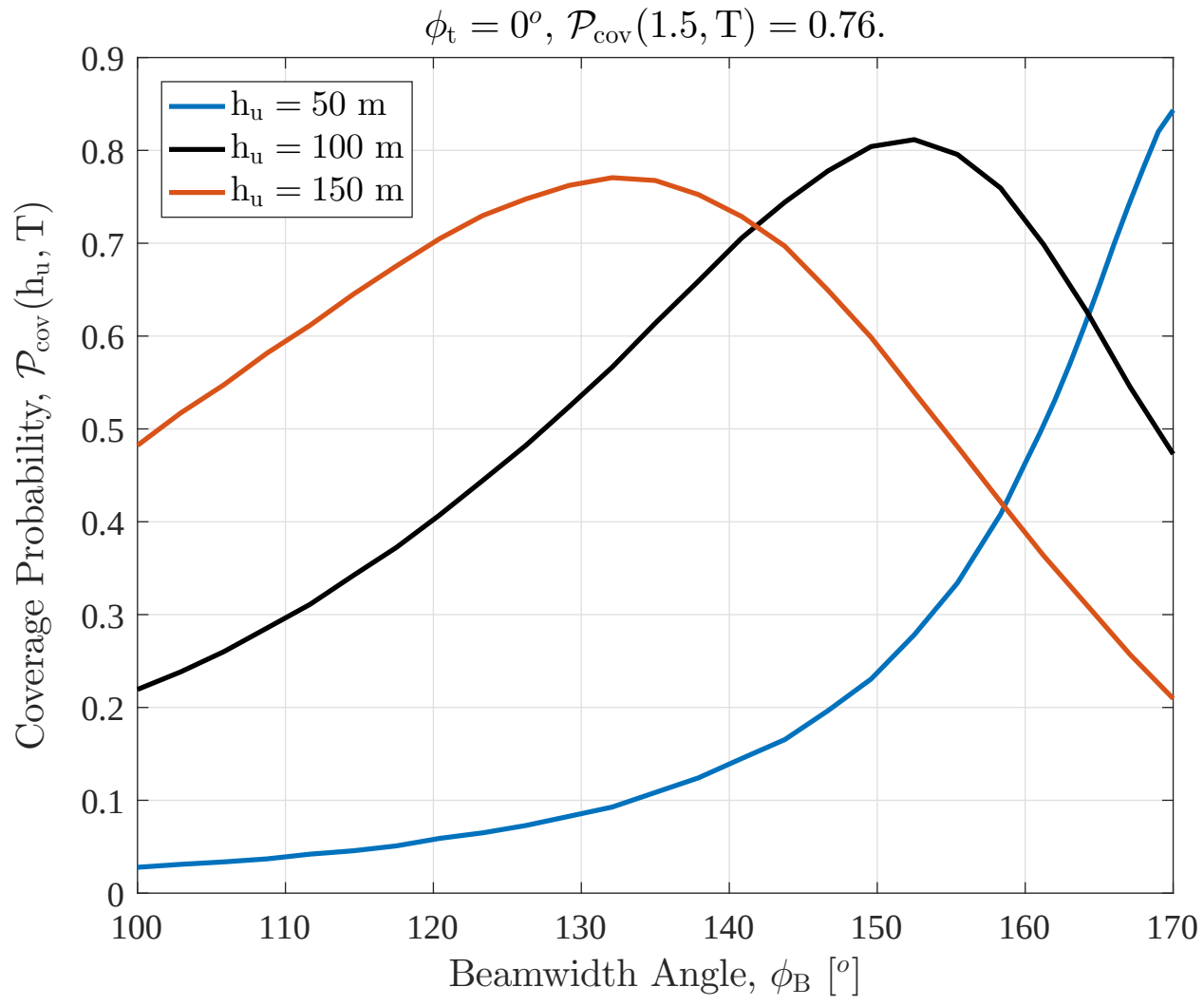

(a)

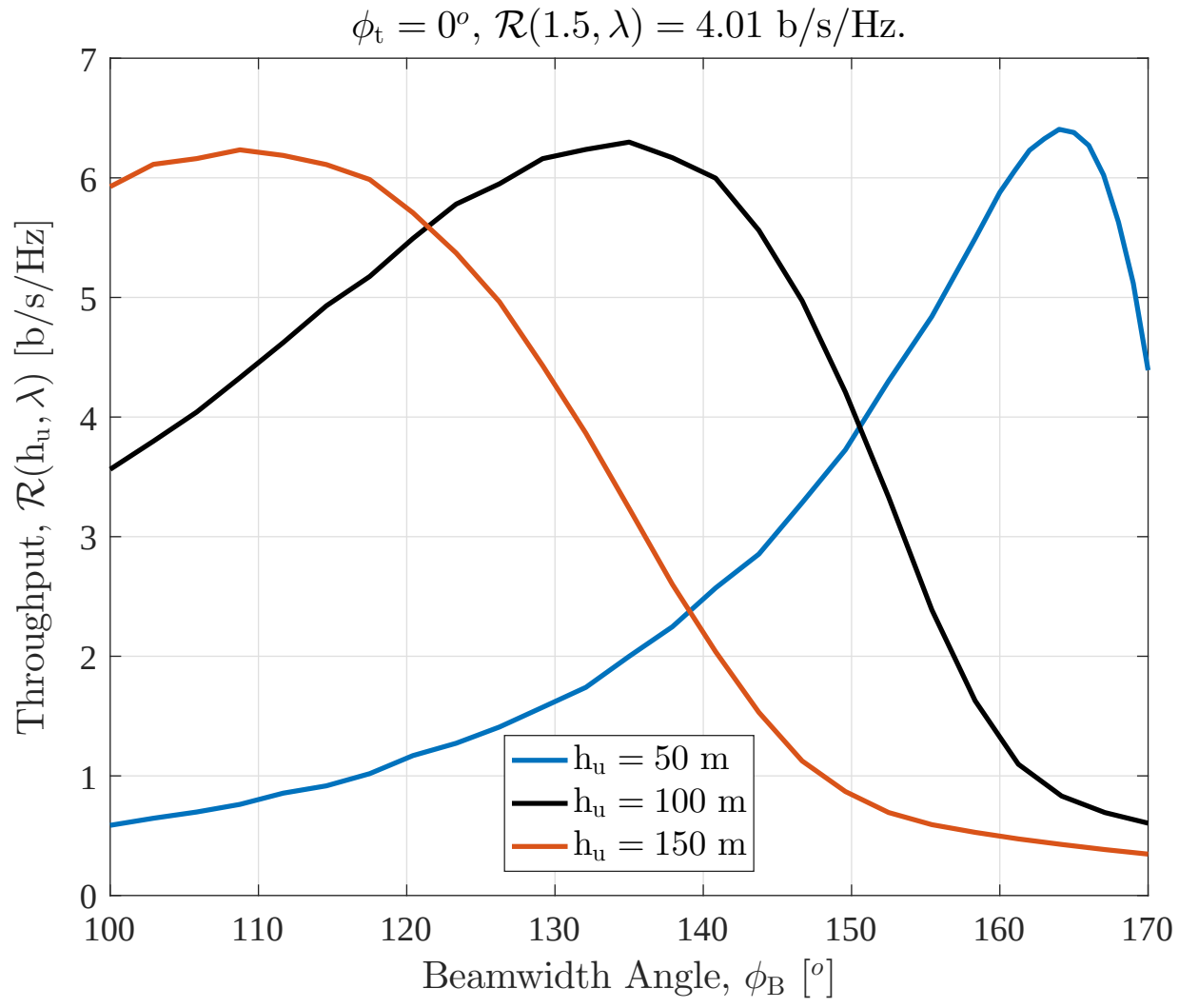

(b)

Fig. 5. Appropriate dimensioning of UAV antenna beamwidth $\phi_\mathrm{B}$ considerably improves the link performance and extends the range of reliable and efficient UAV operation. For each of the performance metrics an optimum $\phi_\mathrm{B}$ exists which depends on the altitude of UAV $\mathrm{h_u}$.

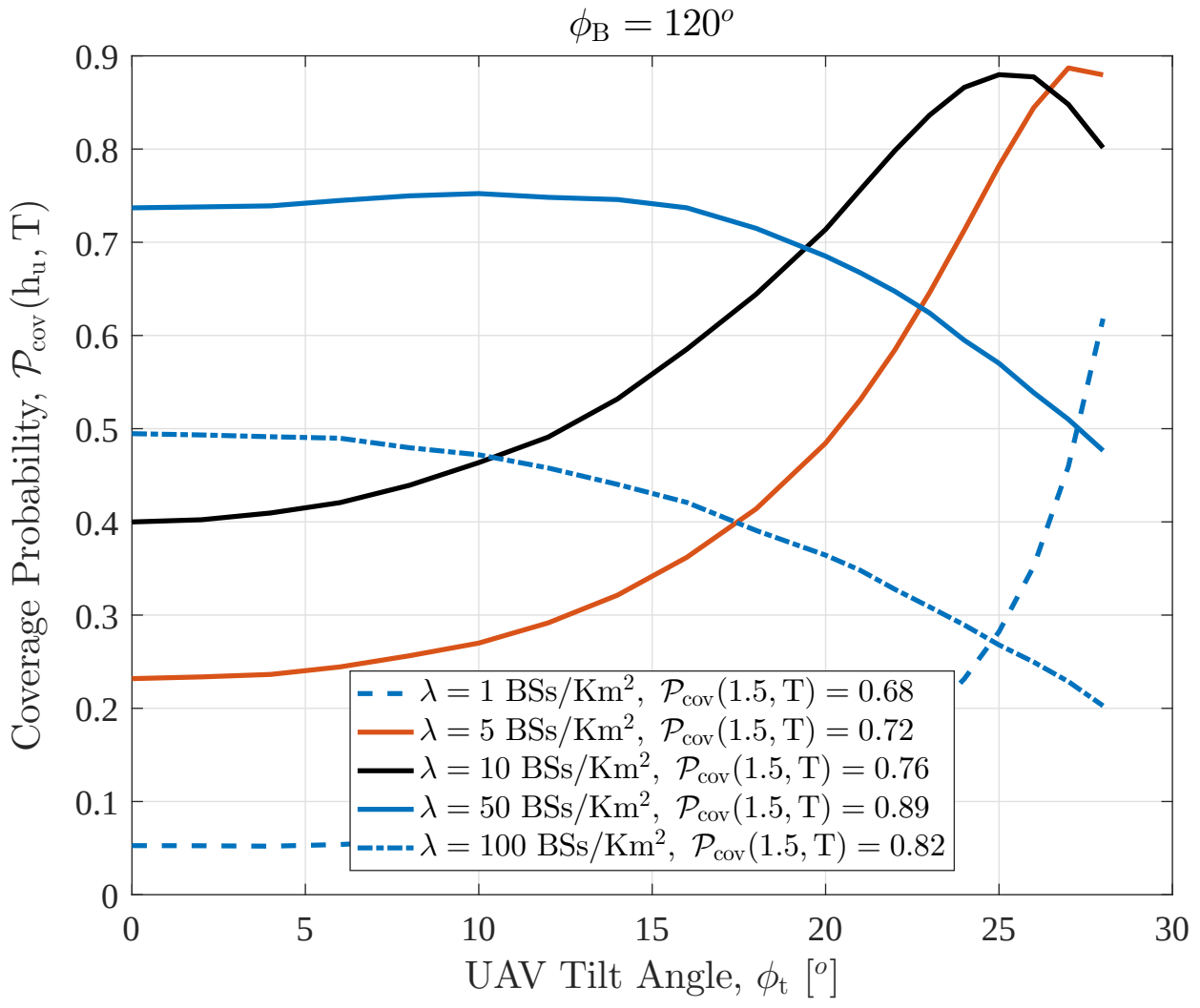

(a)

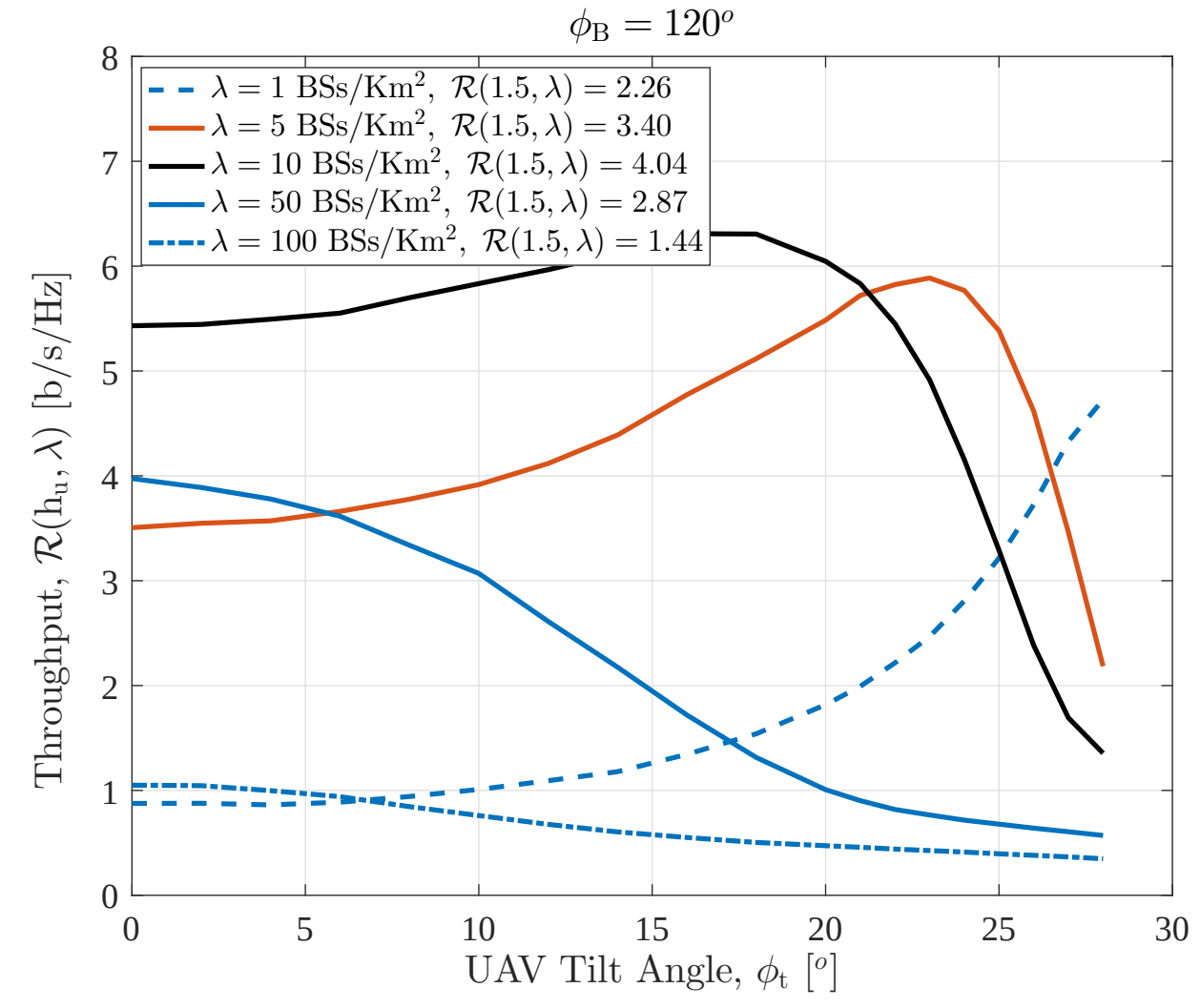

(b)

Fig. 6. A significant performance enhancement is observed by leveraging UAV antenna tilt for sparse to medium dense networks. However, tilting is not beneficial for a dense network due to the inclusion of significantly more interfering BSs.

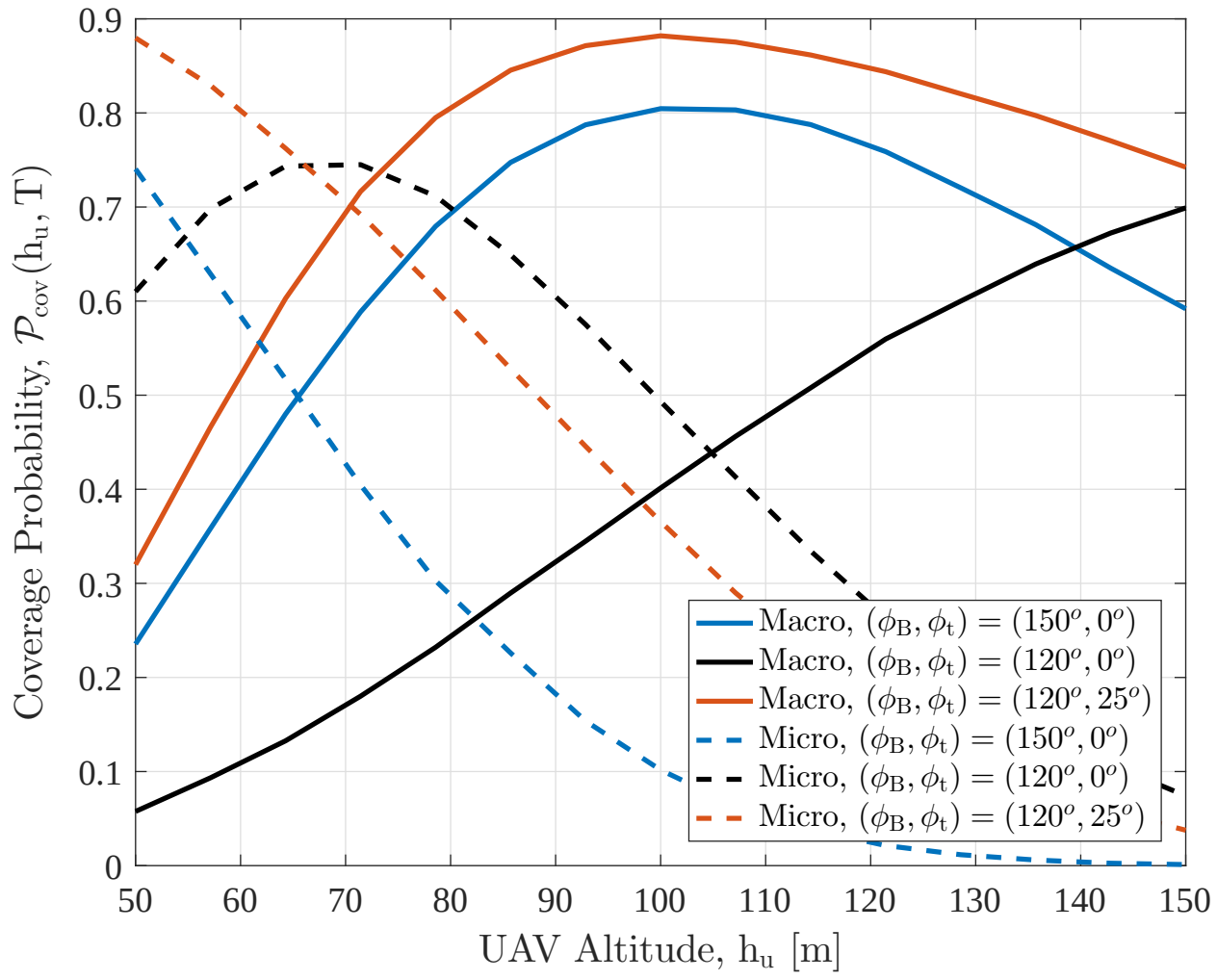

(a)

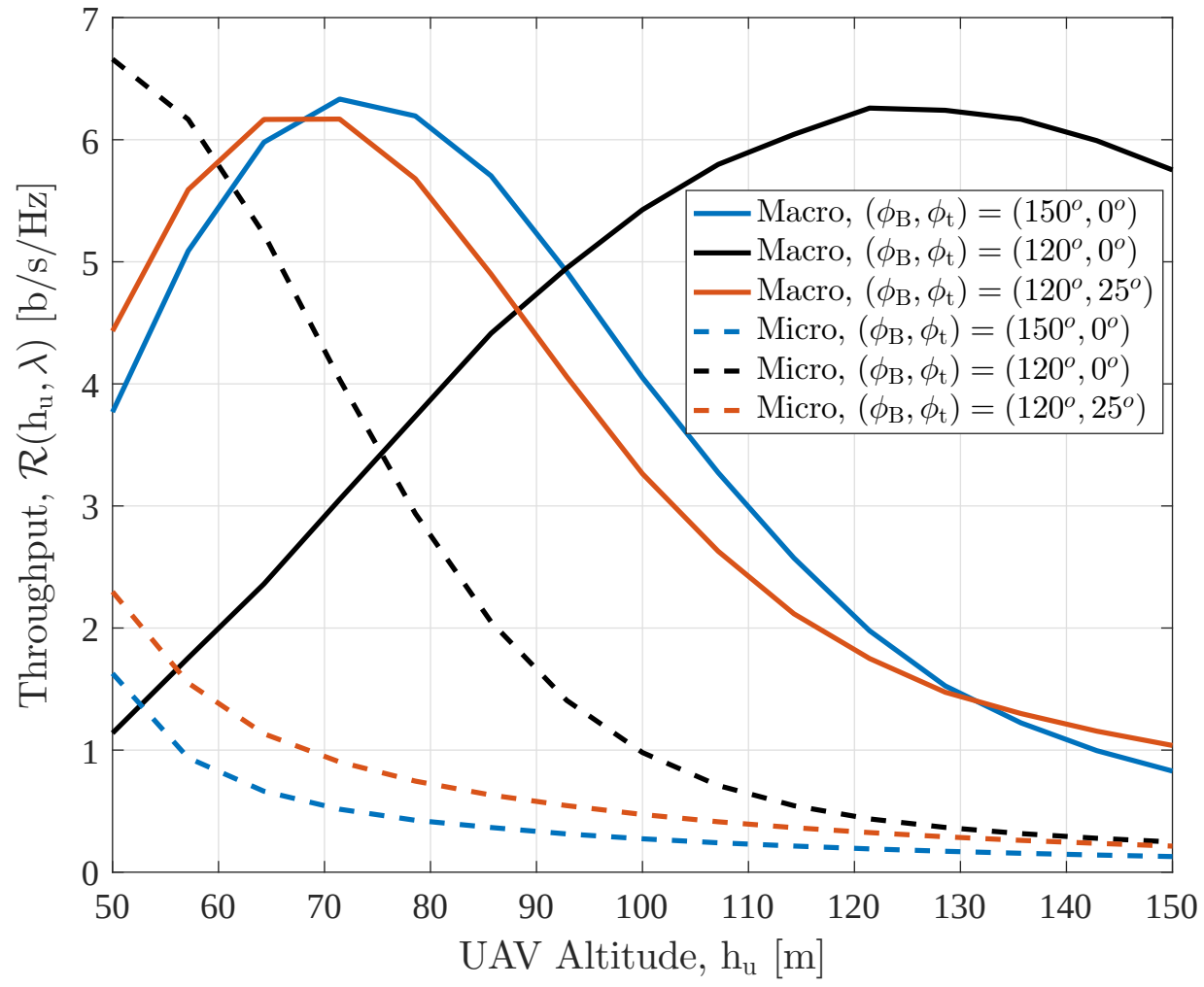

(b)

Fig. 7. At high altitudes macro cells are superior to micro cells for UAV connectivity. In particular there is a certain altitude at which UAV should switch from micro to macro tier. Therefore, an altitude dependent tier connectivity should be taken into account. The cross-over point however depends on the desired performance metric and UAV antenna beamwidth and tilt angle.

than ground users.

*3) **Impact of UAV Beamwidth** :* Results show that an adequate choice of UAV antenna beamwidth $\phi_{\rm B}$ can significantly enhance the network performance, as shown in Figure 5. An optimal antenna beanwidth exists as a higher UAV beamwidth $\phi_{\rm B}$ will increase the probability of seeing at least one BS, hence increasing the coverage probability, but this higher beamwidth also increases the aggregate interference hence lowering the link performance.

Interestingly, the optimal beamwidth that maximizes the coverage probability does not maximizes the throughput. The reason is that throughput is a mean value over whole SINR distribution while the coverage probability depends only on an specific SINR threshold. An optimal design needs hence to choose a compromise within this two conflicting goals, establishing a *throughput-coverage trade-off*.

When using the optimum $\phi_{\rm B}$ a flying UAV can establish a link with even higher capacity and more reliable than the ones established by ground UE with omnidirectional antenna, which enables UAV to safely and efficiently operate at higher altitudes. These figures also show that as $\phi_{\rm B}$ increases, the optimum altitude decreases. This trend holds both for throughput and coverage probability.

*4) **Impact of UAV Antenna Tilt**:* By increasing the tilt angle of the UAV antenna two effects happen simultaneously: the number of BSs within the mainlobe increases, while most of the BSs transit from LoS to NLoS condition. The former effect is constructive due to the presence of more candidate BSs for serving UAV and is destructive because of increased number of interfering BSs. The latter effect is also beneficial as it reduces the LoS interfering BSs and is detrimental as the serving BS might end up in an NLoS condition. These opposite influences eventually result in an optimum tilt angle $\phi_\mathrm{t}$ that in many cases is non zero, as revealed in Figure 6.

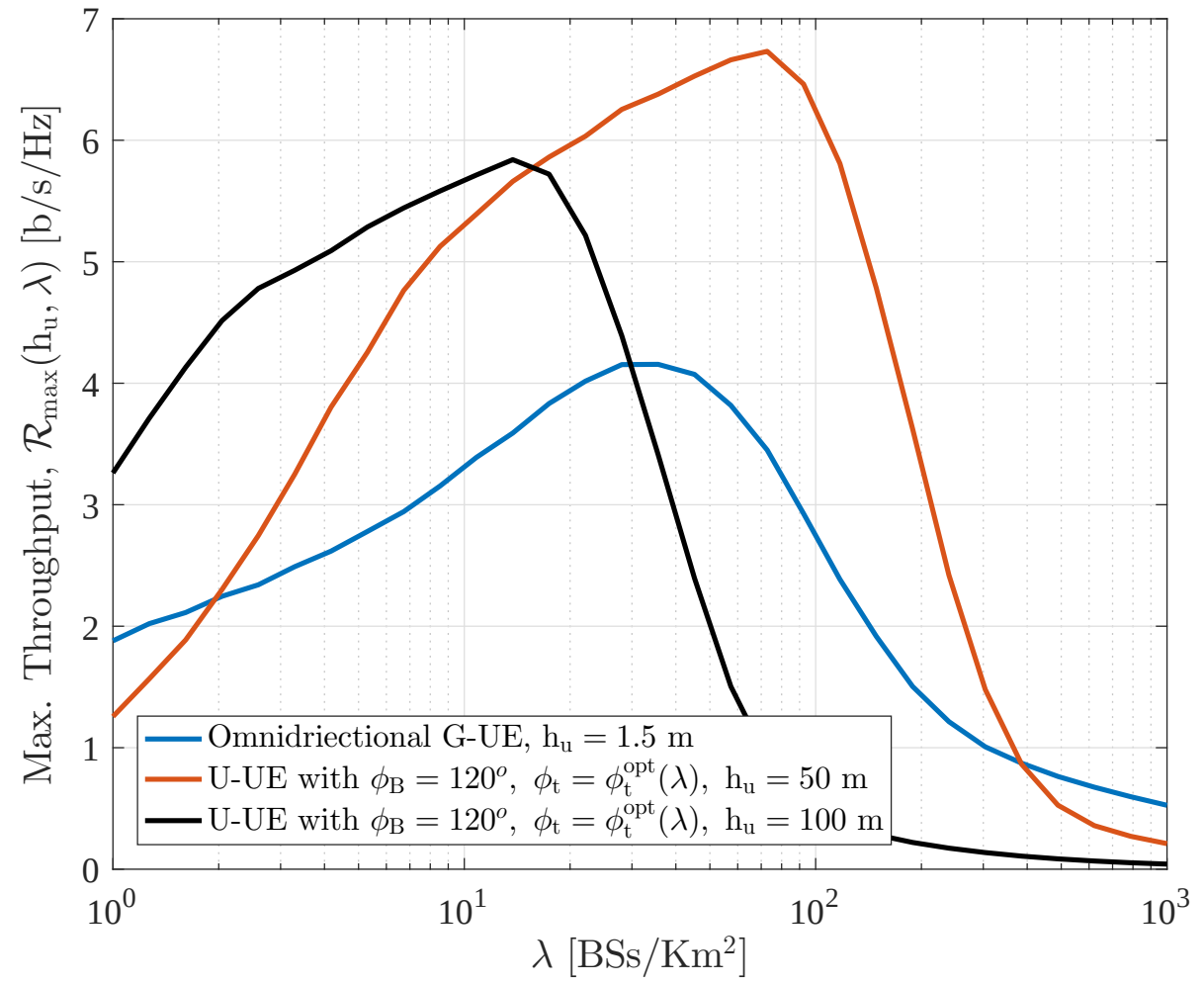

Fig. 8. From UAV perspective network performance converges to zero as it goes dense even though optimal titling angle is adopted. However, lowering UAV altitude partially save the UAV. Note that for these simulations ground UE are considered to be equipped with omnidirectional antenna.

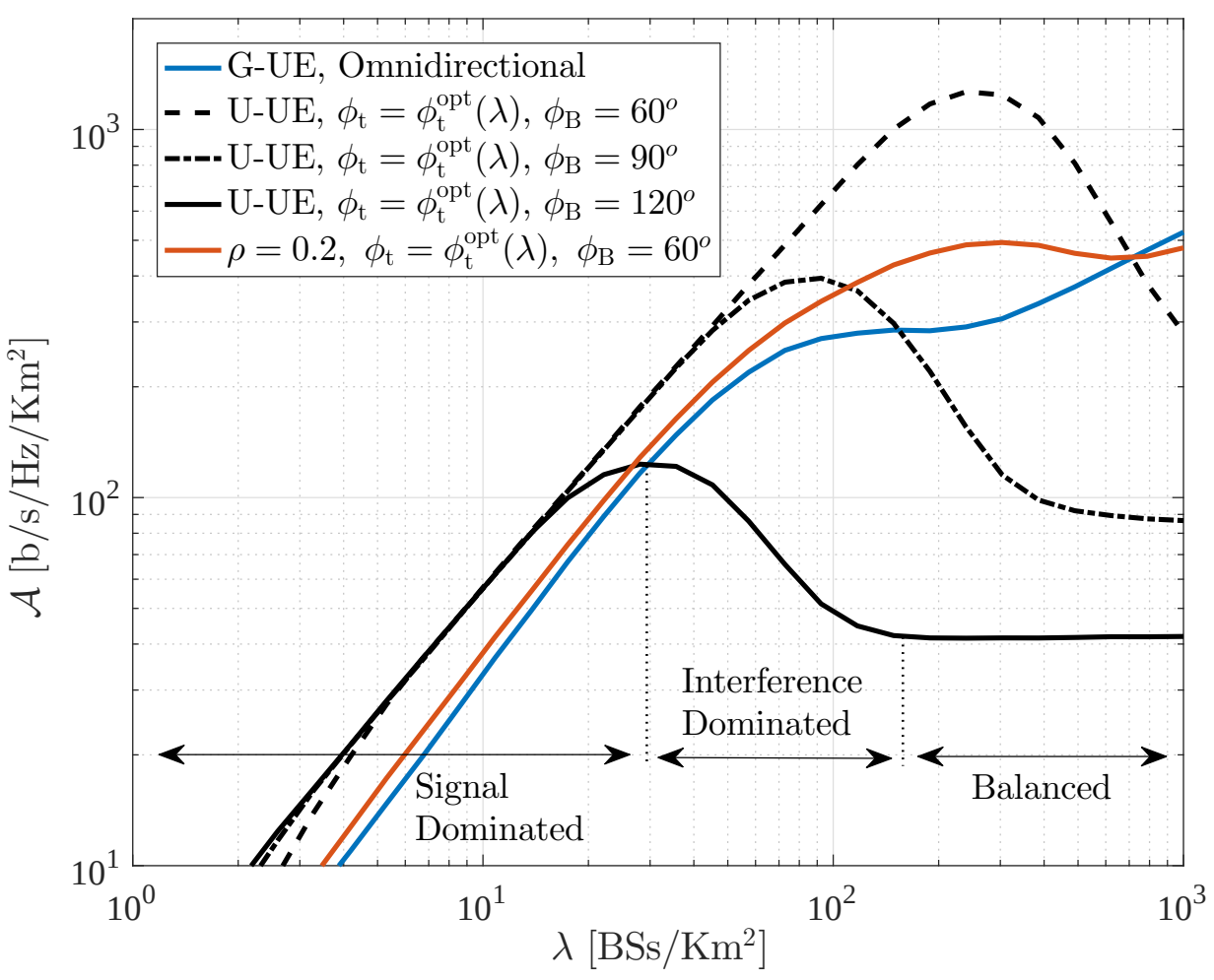

Fig. 9. Adding UAV UEs can increase area spectral efficiency up to a certain density of BSs $\lambda$. Three scaling regimes with respect to $\lambda$ are recognized for aerial users.

Using this optimal tilt considerably increases the quality of service. For instance, in one case when $\lambda = 5$ BSs/Km$^2$ link coverage increases from 23% to 89% and channel capacity is enhanced from 3.5 b/s/Hz to 5.8 b/s/Hz. However, as the network densifies, the optimum $\phi_\mathrm{t}$ converges to zero, as shown in Figure 6. Therefore, tilting the antenna is a good strategy only in non-dense networks.

### C. Heterogeneous Networks - Tier Selection

As macro base station deployment alone does not satisfy the increasing data demand in cellular networks, special attention has been paid to deploying less expensive micro base stations [32]. In general, micro BSs adopt different characteristics including lower transmit power, smaller cell size, lower BS height and larger deployment densities. Moreover, macro and micro BSs can operate on orthogonal frequency spectrum [32] recommended by 3GPP for the design of 4G LTE networks. As a UAV is potentially capable of seeing both tiers in a heterogeneous network and hence receiving their radiated signals, one might ask which tier could be the best candidate for UAV connectivity.

To address this issue, following characteristics of macro and micro BSs as described in [31] including their values of BSs density $\lambda$, height $\mathrm{h_b}$ and transmit power $\mathrm{P_{tx}}$, we use our model to study UAV connectivity performance for different network types (see Figure 7). Intuitively, the lower density of macro BSs than micro BSs imposes less interference and hence is favorable for increasing the UAV performance. However, the higher macro BS generate more interfering LoS BSs, degrading the SINR level. This non-trivial trade-off is studied in Figure 7, which shows that there is a certain altitude where a UAV should switch to macro BSs to improve its performance. Therefore, a smart *altitude aware tier selection strategy* could enhance the coverage and throughput of the UAV communication link.

The switching altitude is lower for larger $\phi_\mathrm{B}$ and also depends on the tilting angle $\phi_\mathrm{t}$. For example, the coverage probability increases from 10% to 80% for a UAV flying at $100\,\mathrm{m}$ with $(\phi_\mathrm{B}, \phi_\mathrm{t}) = (150^o, 0^o)$. A similar conclusion is drawn for throughput evaluation for which the switching altitude is different. For instance, the same case study shows 15 times throughput enhancement by selecting appropriate tier at $\mathrm{h_u} = 100\,\mathrm{m}$. These results confirm that, for the considered target rate $\mathrm{R_t} = 100\,\mathrm{Kbps}$, the height that maximizes the coverage probability is higher than the one that maximizes throughput.

### D. Network Densification - Optimal Density

Network densification is a promising solution in cellular technology to enhance network capacity and user performance through spatial reuse of spectrum. However, this is not always beneficial for the performance of individual UEs, as the individual performance can go to zero even when the global performance grows [8]. It is hence important to determine the optimal network density for users, beyond which further densification is useless or even harmful, particularly when aerial users are added. This operating point depends on several factors such as the altitude for corresponding UEs, the percentage of UAV users and their antenna design.

To study the above-mentioned issue, we first investigate the impact of densification on user performance. For this, for a given UAV user we consider an optimal tuning of $\phi_\mathrm{t}$ at each density denoted by $\phi_\mathrm{t}^{\mathrm{opt}}(\lambda)$, achieving the maximum throughput over all possible values of $\phi_\mathrm{t}$, which is denoted by $\mathcal{R}_{\max}$.

Figure 8 shows that interestingly aerial users are capable of receiving remarkably higher throughput than ground users. However, further increasing $\lambda$ results in a larger loss for UAVs. Moreover, at each altitude user throughput is maximized at a finite density and then decreases to zero as the network is further densified. Comparing two cases for UAV users shows that, in general, *as the network densifies a lower UAV altitude* *leads to significant performance enhancement.* Furthermore, the optimum density for maximum performance decreases as the altitude of UAVs increases.

We also studied the scaling properties, with respect to the BS density, of networks composed solely of ground users, aerial users, and mixed networks. Three scaling regimes are recognized for networks of only aerial users, as illustrated in Figure 9. First, *signal dominated regime* where the signal power growth is dominated over increased interference. In this region, aerial users inclusion with appropriate $\phi_{\mathrm{t}}$ enhances the total ASE, *i.e.* adding UAVs to the network and sharing resources with them overall increases network capacity. Secondly, *interference dominated regime* where the interference growth generated by the BSs dominates and, hence, degrades the overall performance. The effect is significant, and can result in a smaller ASE of UAV networks compared to networks of only ground UEs (when the black lines cross the blue line in Fig. 8). Finally, *balanced regime* where the user achievable throughput is inversely proportional to $\lambda$ resulting in a constant ASE. Therefore, for very dense networks adding more aerial users is detrimental for the network performance. In general, when balancing requirements for ground and aerial users, the optimal network density depends on UAV beamwidth as well as resource requirements $\rho$ of ground and aerial users.

## V. Conclusion

We studied the integration of UAVs as aerial users into current and future cellular networks. For this, a generic framework was developed that allowed us to find closed-form expressions of user and network performance metrics. Design insights were found through the analysis, which might serve to improve UAVs connectivity in cellular networks.

When integrated into the cellular network, our results showed that omnidirectional UAVs are highly vulnerable to interference caused by BSs with which they have line of sight condition. Fortunately, there is still hope for UAV integration, as a careful design of the UAV antenna and other system parameters can greatly improve the UAV performance. In particular, an environment dependent optimum choice of UAV antenna tilt, beamwidth and altitude significantly improve the link coverage probability and achievable throughput. Moreover, heterogeneous networks introduce an additional degree of freedom: UAVs at relatively low altitudes are best served by micro cells, while at higher altitudes macro cells give the best service. An optimal switching altitude can be derived, which depends on the performance requirement — coverage probability or achievable throughput.

Finally, the scaling properties of a network with respect to the BS density were analyzed for both ground and aerial users in terms of spectral efficiency per square meter. According to our results, introducing UAVs can be beneficial for the network performance at moderate BSs density. However, as the BS density grows the benefit of UAVs inclusion decreases. In fact, ultra dense networks perform better when serving only ground devices, and hence adding UAVs is detrimental for the network performance. This suggests that, although the integration of UAVs to the cellular network seems feasible and even desirable, they might decrease network densification gains of operators.

## Appendix A
## Proof of Theorem 1

Using the definition presented in (7), a direct calculation shows that

$$\mathcal{P}_{\mathrm{cov}} = \sum_{\upsilon\in\{\mathrm{L},\mathrm{N}\}} \int_{\mathcal{C}} \mathcal{P}^{\upsilon}_{\mathrm{cov}|R_{\mathrm{S}}}(\mathrm{r_S})\ f^{\upsilon}_{R_{\mathrm{S}}}(\mathrm{r_S})\ \mathrm{r_S}\mathrm{d}\varphi\mathrm{d}\mathrm{r_S}. \tag{31}$$

Above,

$$\mathcal{P}^{\mathrm{L}}_{\mathrm{cov}|R_{\mathrm{S}}} = \mathbb{P}[\mathsf{SINR} > \mathrm{T}|\, R_{\mathrm{S}} = \mathrm{r_S},\, \mathrm{LoS}], \tag{32}$$

$$\mathcal{P}^{\mathrm{N}}_{\mathrm{cov}|R_{\mathrm{S}}} = \mathbb{P}[\mathsf{SINR} > \mathrm{T}|\, R_{\mathrm{S}} = \mathrm{r_S},\, \mathrm{NLoS}], \tag{33}$$

are the conditional coverage probabilities when the serving BS location is given at specific ground location in $\mathbb{R}^2$ and the serving BS is LoS and NLoS respectively. Note that $\mathcal{P}^{\upsilon}_{\mathrm{cov}|R_{\mathrm{S}}}$ is independent of the angular coordinate of the serving BS location in polar coordinate, being only dependent on the ground distance $\mathrm{r_S}$. Also, $f^{\mathrm{L}}_{R_{\mathrm{S}}}(\mathrm{r_S})$ and $f^{\mathrm{N}}_{R_{\mathrm{S}}}(\mathrm{r_S})$ are probability distributions over $\mathrm{r_S}$ that correspond to when the serving BS is in LoS and NLoS condition, respectively. Regardless of the serving link being in LoS or NLoS condition, the received signal is interfered by both LoS and NLoS BSs as stated in (5). Following well-known PPP properties, the function $f^{\mathrm{L}}_{R_{\mathrm{S}}}(\mathrm{r_S})$ can be expressed as

$$f^{\mathrm{L}}_{R_{\mathrm{S}}}(\mathrm{r_S}) = \lambda_{\mathrm{L}}(\mathrm{r_S}) \cdot \mathrm{P}^{\mathrm{L}}_{\mathrm{noL}}(\mathrm{r_S}) \cdot \mathrm{P}^{\mathrm{L}}_{\mathrm{noN}}(\mathrm{r_S}), \tag{34}$$

where $\lambda_{\mathrm{L}}(\mathrm{r_S})$ is the unconditional PDF of having an LoS BS, $\mathrm{P}^{\mathrm{L}}_{\mathrm{noL}}(\mathrm{r_S})$ is the probability of having no LoS BS that provides stronger signal for the UE, and $\mathrm{P}^{\mathrm{L}}_{\mathrm{noN}}(\mathrm{r_S})$ is the probability of having no NLoS BS with better link. Assuming that $\mathcal{A}^{\mathrm{L}}_{\mathrm{noL}}(\mathrm{r_S})$ is formed by all the 2D locations at which a LoS BS could provide a better link, $\mathrm{P}^{\mathrm{L}}_{\mathrm{noL}}(\mathrm{r_S})$ can be written as

$$\mathrm{P}^{\mathrm{L}}_{\mathrm{noL}}(\mathrm{r_S}) = e^{-2\int_{\mathcal{A}^{\mathrm{L}}_{\mathrm{noL}}} \lambda_{\mathrm{L}}(r)\ r\mathrm{d}\varphi\mathrm{d}r}. \tag{35}$$

Similarly, if $\mathcal{A}^{\mathrm{L}}_{\mathrm{noN}}(\mathrm{r_S})$ is defined as the set of locations with stronger NLoS signal than the LoS signal at the location $\mathrm{r_S}$, one finds that

$$\mathrm{P}^{\mathrm{L}}_{\mathrm{noN}}(\mathrm{r_S}) = e^{-2\int_{\mathcal{A}^{\mathrm{L}}_{\mathrm{noN}}} \lambda_{\mathrm{N}}(r)\ r\mathrm{d}\varphi\mathrm{d}r}. \tag{36}$$

The sets $\mathcal{A}^{\mathrm{L}}_{\mathrm{noL}}$ and $\mathcal{A}^{\mathrm{L}}_{\mathrm{noN}}$ are dependent on the geometry of the network and are characterized in the sequel. For this, first we note that the area $\mathcal{C}$ covered by a UAV of beamwidth angle $\phi_{\mathrm{B}}$ and tilt angle $\phi_{\mathrm{t}}$ is an elliptical section characterized by its semi-major and semi-minor axis, i.e. $\mathrm{r_M}$ and $\mathrm{r_m}$ respectively, and its origin $(\mathrm{r_e}, 0)$ as illustrated in Figure 10. Following [33] we can obtain these parameters in Table III where we use the following notations

$$[y]^{+}_{x} \triangleq \max(x, y), \quad [y]^{-}_{x} \triangleq \min(x, y). \tag{37}$$

The set $\mathcal{A}^{\mathrm{L}}_{\mathrm{noL}}$ can be written as the intersection of $\mathcal{C}$ and a disc of radius $\mathrm{r_S}$ centered at origin as illustrated in Figure 10. Therefore, one can obtain

$$\mathcal{A}^{\mathrm{L}}_{\mathrm{noL}} = \Big\{(r,\varphi)\ \Big|\ r < \mathrm{r}^{\mathrm{L}}_{\mathrm{L}},\ \varphi \in [0,\pi]\backslash[\varphi_1(r), \varphi_2(r)]\Big\}, \tag{38}$$

where $\mathrm{r}^{\mathrm{L}}_{\mathrm{L}} = \mathrm{r_S}$ and by using geometry properties, $\varphi_1(r)$ and $\varphi_2(r)$ are obtained in Table III.

TABLE III. Parameters values.

| |
|---|
| $\mathrm{r_M} = \frac{\Delta_\mathrm{h}\sin(\phi_\mathrm{B})}{2[\cos^2(\phi_\mathrm{t})-\sin^2(\phi_\mathrm{B}/2)]}$ |
| $\mathrm{r_m} = \frac{\Delta_\mathrm{h}\sin(\phi_\mathrm{B}/2)}{\sqrt{\cos^2(\phi_\mathrm{t})-\sin^2(\phi_\mathrm{B}/2)}}$ |
| $\mathrm{r_e} = \Delta_\mathrm{h}\tan(\phi_\mathrm{t} - \frac{\phi_\mathrm{B}}{2}) + \mathrm{r_M}$ |
| $\mathrm{r_0} = [\mathrm{r_e} - \mathrm{r_M}]_0^+$ |
| $r_\mathrm{L}^\mathrm{L} = \mathrm{r_S},\ \ r_\mathrm{N}^\mathrm{L} = \sqrt{\Big[(\mathrm{A_N}/\mathrm{A_L})^{2/\alpha_\mathrm{N}}\,(\mathrm{r_S^2}+\Delta_\mathrm{h}^2)^{\alpha_\mathrm{L}/\alpha_\mathrm{N}} - \Delta_\mathrm{h}^2\Big]_{\mathrm{r_0^2}}^+}$ |
| $r_\mathrm{L}^\mathrm{N} = \Big[\sqrt{(\mathrm{A_L}/\mathrm{A_N})^{2/\alpha_\mathrm{L}}\,(\mathrm{r_S^2}+\Delta_\mathrm{h}^2)^{\alpha_\mathrm{N}/\alpha_\mathrm{L}} - \Delta_\mathrm{h}^2}\Big]_{\mathrm{r_e}+\mathrm{r_M}}^-,\ \ r_\mathrm{N}^\mathrm{N} = \mathrm{r_S}$ |
| $\varphi_1(r) = \varphi_2(r) = \frac{\pi}{2};\quad \text{if } r < \frac{\mathrm{r_m}}{\mathrm{r_M}}\sqrt{\mathrm{r_M^2}-\mathrm{r_e^2}}\ \&\ \phi_\mathrm{t} < \frac{\phi_\mathrm{B}}{2}$ |
| $\varphi_1(r) = \cos^{-1}\left[\frac{\mathrm{r_e r_m^2} - \sqrt{\mathrm{r_e^2 r_m^4} - (\mathrm{r_m^2}-\mathrm{r_M^2})(\mathrm{r_e^2 r_m^2} + r^2\mathrm{r_M^2} - \mathrm{r_m^2 r_M^2})}}{r(\mathrm{r_m^2}-\mathrm{r_M^2})}\right]$, $\varphi_2(r) = \cos^{-1}\left[\frac{\mathrm{r_e r_m^2} + \sqrt{\mathrm{r_e^2 r_m^4} - (\mathrm{r_m^2}-\mathrm{r_M^2})(\mathrm{r_e^2 r_m^2} + r^2\mathrm{r_M^2} - \mathrm{r_m^2 r_M^2})}}{r(\mathrm{r_m^2}-\mathrm{r_M^2})}\right]$; if $\frac{\mathrm{r_m}}{\mathrm{r_M}}\sqrt{\mathrm{r_M^2}-\mathrm{r_e^2}} \le r \le \mathrm{r_M} - \mathrm{r_e}\ \&\ \phi_\mathrm{t} < \frac{\phi_\mathrm{B}}{2}$ |
| $\varphi_1(r) = \cos^{-1}\left[\frac{\mathrm{r_e r_m^2} - \sqrt{\mathrm{r_e^2 r_m^4} - (\mathrm{r_m^2}-\mathrm{r_M^2})(\mathrm{r_e^2 r_m^2} + r^2\mathrm{r_M^2} - \mathrm{r_m^2 r_M^2})}}{r(\mathrm{r_m^2}-\mathrm{r_M^2})}\right]$, $\varphi_2(r) = \pi;\ \text{if } r > \lvert\mathrm{r_M} - \mathrm{r_e}\rvert$ |

Therefore, the integral in (35) can be written as

$$
\begin{aligned}
\int_{\mathcal{A}_\mathrm{noL}^\mathrm{L}} \lambda_\mathrm{L}(r)\ r\mathrm{d}\varphi\mathrm{d}r &= \int_0^{\mathrm{r_S}}\int_0^{\varphi_1(r)} \lambda\mathcal{P}_\mathrm{L}(r)\ r\mathrm{d}\varphi\mathrm{d}r \\
&+ \int_0^{\mathrm{r_S}}\int_{\varphi_2(r)}^{\pi} \lambda\mathcal{P}_\mathrm{L}(r)\ r\mathrm{d}\varphi\mathrm{d}r \\
&= \lambda\int_0^{\mathrm{r_S}} r\varphi_1(r)\mathcal{P}_\mathrm{L}(r)\ \mathrm{d}r \\
&+ \lambda\int_0^{\mathrm{r_S}} r[\pi - \varphi_2(r)]\mathcal{P}_\mathrm{L}(r)\ \mathrm{d}r \\
&= \lambda\int_0^{\mathrm{r_S}} r\mathcal{P}_\mathrm{L}(r)[\varphi_1(r) + \pi - \varphi_2(r)]\ \mathrm{d}r \\
&\triangleq \lambda\mathcal{I}_\mathrm{1L}^\mathrm{L}
\end{aligned} \tag{39}
$$

The set $\mathcal{A}_\mathrm{noN}^\mathrm{L}$ can be written as

$$
\mathcal{A}_\mathrm{noN}^\mathrm{L} = \Big\{(r,\varphi)\ \Big|\ r < \mathrm{r_N^L},\ \varphi\in[0,\pi]\backslash[\varphi_1(r),\varphi_2(r)]\Big\}, \tag{40}
$$

where $\mathrm{r_N^L}$ is obtained in Table III, and hence

$$
\begin{aligned}
\int_{\mathcal{A}_\mathrm{noN}^\mathrm{L}} \lambda_\mathrm{N}(r)\ r\mathrm{d}\varphi\mathrm{d}r &= \lambda\int_0^{\mathrm{r_N^L}} r\varphi_1(r)\mathcal{P}_\mathrm{N}(r)\ \mathrm{d}r \\
&+ \lambda\int_0^{\mathrm{r_N^L}} r[\pi - \varphi_2(r)]\mathcal{P}_\mathrm{N}(r)\ \mathrm{d}r \\
&= \lambda\int_0^{\mathrm{r_N^L}} r\mathcal{P}_\mathrm{N}(r)[\varphi_1(r) + \pi - \varphi_2(r)]\ \mathrm{d}r \\
&\triangleq \lambda\mathcal{I}_\mathrm{1N}^\mathrm{L}
\end{aligned} \tag{41}
$$

Using (34) – (41), $f_{\mathrm{R_S}}^\mathrm{L}(\mathrm{r_S})$ can be finally written as

$$
f_{\mathrm{R_S}}^\mathrm{L}(\mathrm{r_S}) = \lambda\mathcal{P}_\mathrm{L}(\mathrm{r_S})\cdot e^{-2\lambda\left[\mathcal{I}_\mathrm{1L}^\mathrm{L} + \mathcal{I}_\mathrm{1N}^\mathrm{L}\right]}. \tag{42}
$$

Similarly we can write

$$
f_{\mathrm{R_S}}^\mathrm{N}(\mathrm{r_S}) = \lambda\mathcal{P}_\mathrm{N}(\mathrm{r_S})\cdot e^{-2\lambda\left[\mathcal{I}_\mathrm{1L}^\mathrm{N} + \mathcal{I}_\mathrm{1N}^\mathrm{N}\right]}, \tag{43}
$$

where the integrals and corresponding parameters are defined in Theorem 1.

To obtain the conditional coverage probability $\mathcal{P}_{\mathrm{cov}|R_\mathrm{S}}^{\upsilon}$ one can write

$$
\begin{aligned}
\mathcal{P}_{\mathrm{cov}|R_\mathrm{S}}^{\upsilon} &= \mathbb{P}\left[\frac{\mathrm{P_{tx}}\,\mathrm{G_{tot}}\,\zeta_\upsilon\,\Omega_\upsilon}{\mathrm{N_0}+I} > \mathrm{T}\ \Big|\ R_\mathrm{S} = \mathrm{r_S}\right] \\
&= \mathbb{E}_I\left\{\mathbb{P}\left[\Omega_\upsilon > \frac{\mathrm{T}}{\mathrm{P_{tx}}\,\mathrm{G_{tot}}\,\zeta_\upsilon}(\mathrm{N_0}+I)\ \Big|\ R_\mathrm{S} = \mathrm{r_S}\right]\right\} \\
&\overset{(a)}{=} \mathbb{E}_I\left\{\sum_{k=0}^{m_\upsilon-1}\frac{y_\upsilon^k}{k!}(\mathrm{N_0}+I)^k\exp[-y_\upsilon(\mathrm{N_0}+I)]\ \Big|\ R_\mathrm{S} = \mathrm{r_S}\right\} \\
&= \mathbb{E}_I\left\{\sum_{k=0}^{m_\upsilon-1}\frac{y_\upsilon^k}{k!}e^{-\mathrm{N_0}y_\upsilon}\sum_{j=0}^{k}\binom{k}{j}\mathrm{N_0^{k-j}}I^j\exp[-y_\upsilon I]\ \Big|\ R_\mathrm{S} = \mathrm{r_S}\right\} \\
&= \sum_{k=0}^{m_\upsilon-1} q_k\cdot\mathbb{E}_I\left\{I^k\exp(-y_\upsilon I)\ \Big|\ R_\mathrm{S} = \mathrm{r_S}\right\} \\
&= \sum_{k=0}^{m_\upsilon-1}(-1)^k q_k\cdot\frac{d^k}{dy_\upsilon^k}\mathcal{L}_{I|R_\mathrm{S}}^{\upsilon}(y_\upsilon),
\end{aligned} \tag{44}
$$

where (a) follows from the gamma distribution of $\Omega_\upsilon$ with an integer parameter $m_\upsilon$ and $q_k$ and $y_\upsilon$ are expressed in (13) and (14) respectively.

To derive $\mathcal{L}_{I|R_\mathrm{S}}(y_\upsilon)$ one can write

$$
\begin{aligned}
\mathcal{L}_{I|X_\mathrm{S}}(y_\upsilon) &= \mathbb{E}_I\left\{e^{-y_\upsilon I}\ \Big|\ R_\mathrm{S} = \mathrm{r_S}\right\} \\
&= \mathbb{E}_{\Phi,\Omega}\left\{\prod_{r\in\Phi\backslash\mathrm{r_S}} e^{-y_\upsilon\mathrm{P_{tx}}\,\mathrm{G_{tot}}\,\zeta_\xi(r)\,\Omega_\xi}\right\} \\
&= \mathbb{E}_{\Phi}\left\{\prod_{r\in\Phi\backslash\mathrm{r_S}}\mathbb{E}_\Omega\left\{e^{-y_\upsilon\mathrm{P_{tx}}\,\mathrm{G_{tot}}\,\zeta_\xi(r)\,\Omega_\xi}\right\}\right\}.
\end{aligned}
$$

The above equation can be further processed as

$$
\begin{aligned}
\mathcal{L}_{I|R_\mathrm{S}}(y_\upsilon) &= \mathbb{E}_{\Phi_\mathrm{L}}\left\{\prod_{r\in\Phi_\mathrm{L}\backslash\mathrm{r_S}}\mathbb{E}_\Omega\left\{e^{-y_\upsilon\mathrm{P_{tx}}\,\mathrm{G_{tot}}\,\zeta_\mathrm{L}(r)\,\Omega_\mathrm{L}}\right\}\right\} \\
&\times\mathbb{E}_{\Phi_\mathrm{N}}\left\{\prod_{r\in\Phi_\mathrm{N}\backslash\mathrm{r_S}}\mathbb{E}_\Omega\left\{e^{-y_\upsilon\mathrm{P_{tx}}\,\mathrm{G_{tot}}\,\zeta_\mathrm{N}(r)\,\Omega_\mathrm{N}}\right\}\right\} \\
&\overset{(a)}{=} e^{-2\int_{\bar{\mathcal{A}}_\mathrm{noL}^\upsilon}\lambda_\mathrm{L}(r)\,[1-\Upsilon_\mathrm{L}(r,y_\upsilon)]\,r\mathrm{d}\varphi\mathrm{d}r} \\
&\times e^{-2\int_{\bar{\mathcal{A}}_\mathrm{noN}^\upsilon}\lambda_\mathrm{N}(r)\,[1-\Upsilon_\mathrm{N}(r,y_\upsilon)]\,r\mathrm{d}\varphi\mathrm{d}r}
\end{aligned}
$$

where (a) is obtained using the probability generating functional (PGFL) of PPP. Moreover, in above equation $\bar{\mathcal{A}}_{\mathrm{no}\xi}^\upsilon$ indicates the complementary of the sets $\mathcal{A}_{\mathrm{no}\xi}^\upsilon$ over $\mathcal{C}$, i.e.

$$
\bar{\mathcal{A}}_{\mathrm{no}\xi}^\upsilon = \mathcal{C}\backslash\mathcal{A}_{\mathrm{no}\xi}^\upsilon. \tag{45}
$$

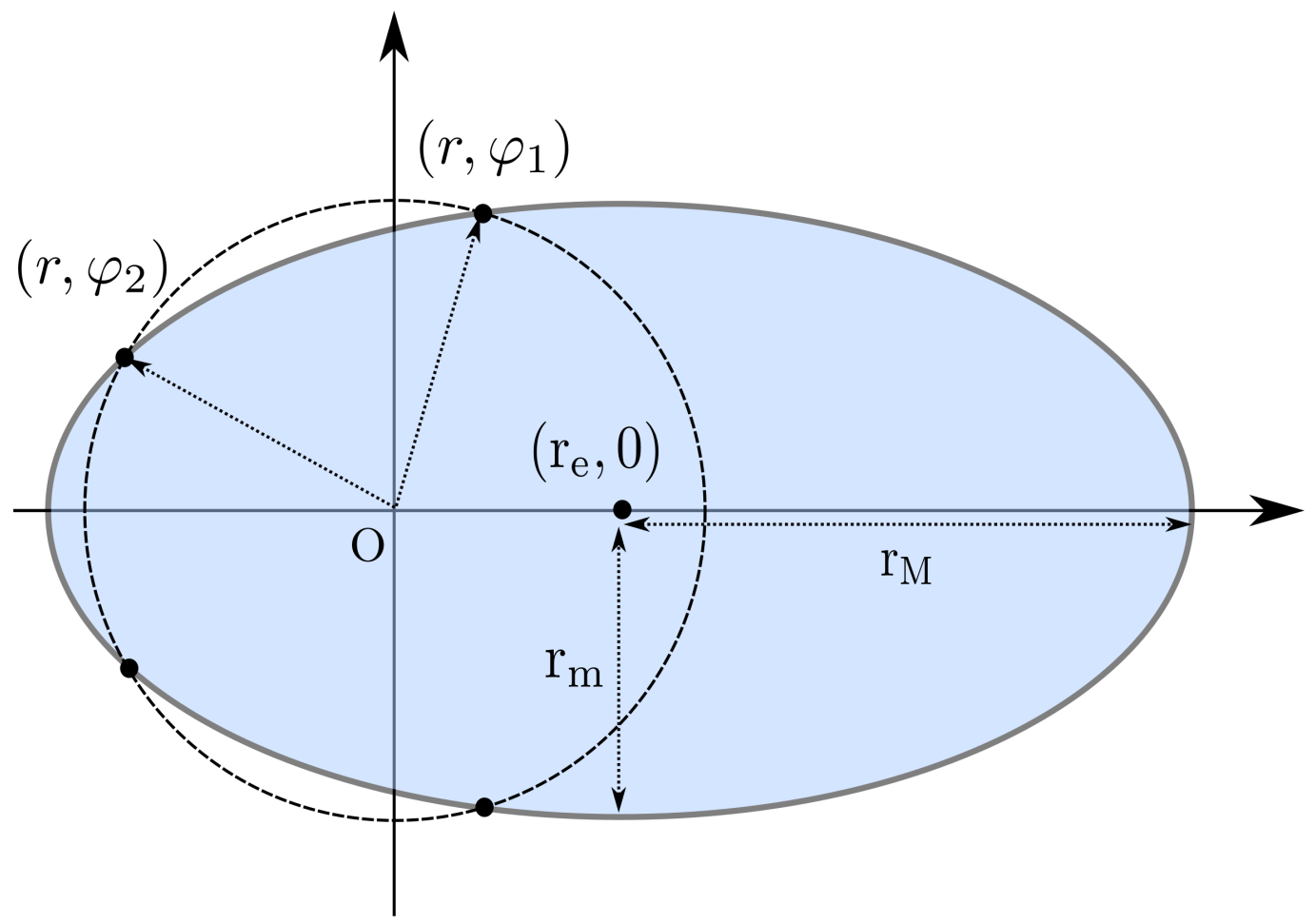

Fig. 10. An illustration of the region $\mathcal{C}$ for the calculation of $\varphi_1$ and $\varphi_2$.

Thus,

$$
\begin{aligned}
&\int_{\bar{\mathcal{A}}^{\mathrm{L}}_{\mathrm{noL}}} \lambda_{\mathrm{L}}(r)\left[1-\Upsilon_{\mathrm{L}}(r,y_{\mathrm{L}})\right]\mathrm{d}x \\
&= \int_{\mathrm{r_S}}^{\mathrm{r_e+r_M}} \int_{0}^{\varphi_1(r)} \lambda_{\mathrm{L}}(r)\left[1-\Upsilon_{\mathrm{L}}(r,y_{v})\right] r\mathrm{d}\varphi\mathrm{d}r \\
&+ \int_{\mathrm{r_S}}^{\mathrm{r_e+r_M}} \int_{\varphi_2(r)}^{\pi} \lambda_{\mathrm{L}}(r)\left[1-\Upsilon_{\mathrm{L}}(r,y_{\mathrm{L}})\right] r\mathrm{d}\varphi\mathrm{d}r \\
&= \lambda \int_{\mathrm{r_S}}^{\mathrm{r_e+r_M}} \varphi_1(r)\mathcal{P}_{\mathrm{L}}(r)\left[1-\Upsilon_{\mathrm{L}}(r,y_{\mathrm{L}})\right] r\mathrm{d}r \\
&+ \lambda \int_{\mathrm{r_S}}^{\mathrm{r_e+r_M}} [\pi-\varphi_2(r)]\mathcal{P}_{\mathrm{L}}(r)\left[1-\Upsilon_{\mathrm{L}}(r,y_{\mathrm{L}})\right] r\mathrm{d}r \\
&= \lambda \int_{\mathrm{r_S}}^{\mathrm{r_e+r_M}} r\mathcal{P}_{\mathrm{L}}(r)\left[1-\Upsilon_{\mathrm{L}}(r,y_{\mathrm{L}})\right][\varphi_1(r)+\pi-\varphi_2(r)]\,\mathrm{d}r \\
&\triangleq \lambda \mathcal{I}^{\mathrm{L}}_{2\mathrm{L}}
\end{aligned} \tag{46}
$$

and

$$
\begin{aligned}
&\int_{\bar{\mathcal{A}}^{\mathrm{L}}_{\mathrm{noN}}} \lambda_{\mathrm{N}}(r)\left[1-\Upsilon_{\mathrm{N}}(r,y_{\mathrm{L}})\right] r\mathrm{d}\varphi\mathrm{d}r \\
&= \lambda \int_{\mathrm{r^L_N}}^{\mathrm{r_e+r_M}} \varphi_1(r)\mathcal{P}_{\mathrm{N}}(r)\left[1-\Upsilon_{\mathrm{N}}(r,y_{\mathrm{L}})\right] r\mathrm{d}r \\
&+ \lambda \int_{\mathrm{r^L_N}}^{\mathrm{r_e+r_M}} [\pi-\varphi_2(r)]\mathcal{P}_{\mathrm{N}}(r)\left[1-\Upsilon_{\mathrm{N}}(r,y_{\mathrm{L}})\right] r\mathrm{d}r \\
&= \lambda \int_{\mathrm{r^L_N}}^{\mathrm{r_e+r_M}} r\mathcal{P}_{\mathrm{N}}(r)\left[1-\Upsilon_{\mathrm{N}}(r,y_{\mathrm{L}})\right][\varphi_1(r)+\pi-\varphi_2(r)]\,\mathrm{d}r \\
&\triangleq \lambda \mathcal{I}^{\mathrm{L}}_{2\mathrm{N}}
\end{aligned} \tag{47}
$$

Therefore

$$
\mathcal{L}_{I|R_{\mathrm{S}}}(y_{\mathrm{L}}) = e^{-2\lambda\left[\mathcal{I}^{\mathrm{L}}_{2\mathrm{L}}+\mathcal{I}^{\mathrm{L}}_{2\mathrm{N}}\right]}. \tag{48}
$$

Similarly

$$
\mathcal{L}_{I|R_{\mathrm{S}}}(y_{\mathrm{N}}) = e^{-2\lambda\left[\mathcal{I}^{\mathrm{N}}_{2\mathrm{L}}+\mathcal{I}^{\mathrm{N}}_{2\mathrm{N}}\right]}, \tag{49}
$$

where corresponding parameters are defined in Theorem 1.

## Appendix B
## Proof of Lemma 1

The first moment of $I$ can be calculated from its Laplace transform as

$$
\mu_{I|R_{\mathrm{S}}} = -\left.\frac{d}{dy_{\mathrm{L}}}\mathcal{L}_{I|R_{\mathrm{S}}}(y_{\mathrm{L}})\right|_{y_{\mathrm{L}}=0}. \tag{50}
$$

Using (26), one can re=write (50) as

$$
\begin{aligned}
\mu_{I|R_{\mathrm{S}}} &= -2\lambda \left.\frac{d}{dy_{\mathrm{L}}}\mathcal{I}^{\mathrm{L}}_{2\mathrm{L}}\right|_{y_{\mathrm{L}}=0} \cdot \mathcal{L}_{I|R_{\mathrm{S}}}(0) \\
&= 2\lambda \int_{\mathrm{r_S}}^{\mathrm{r_e+r_M}} \left.\frac{d}{dy_{\mathrm{L}}}\Upsilon_{\mathrm{L}}(r,y_{\mathrm{L}})\right|_{y_{\mathrm{L}}=0} r\mathcal{P}_{\mathrm{L}}(r)\left[\varphi_1(r)+\pi-\varphi_2(r)\right]\mathrm{d}r \\
&= 2\lambda \mathrm{P_{tx}G_{tot}} \int_{\mathrm{r_S}}^{\mathrm{r_e+r_M}} r\mathcal{P}_{\mathrm{L}}(r)\zeta_{\mathrm{L}}(r)[\pi+\varphi_1(r)-\varphi_2(r)]\,\mathrm{d}r.
\end{aligned} \tag{51}
$$

For calculating the variance of $I$, one can use the following relationship:

$$
\sigma^2_{I|R_{\mathrm{S}}} = \left.\frac{d^2}{dy^2_{\mathrm{L}}}\mathcal{L}_{I|R_{\mathrm{S}}}(y_{\mathrm{L}})\right|_{y_{\mathrm{L}}=0} - \mu^2_{I|R_{\mathrm{S}}}. \tag{52}
$$

By applying (26) and following a similar derivation than the above, one can find the desired result.

## Appendix C
## Proof of Corollary 1

Let us assume $\phi_{\mathrm{t}} = 0$. Then, $\varphi_1(r) = \varphi_2(r) = \pi/2$ holds, and hence the integral in (21) can be written as

$$
\begin{aligned}
\mu_{I|R_{\mathrm{S}}} &= 2\pi\lambda\mathrm{P_{tx}G_{tot}} \sum_{\mathrm{k=i}}^{\mathrm{j}} \mathrm{p_k} \int_{\mathrm{r_k}}^{\mathrm{r_{k+1}}} r\zeta_{\mathrm{L}}(r)\,\mathrm{d}r \\
&= 2\pi\lambda\mathrm{P_{tx}G_{tot}A_L} \sum_{\mathrm{k=i}}^{\mathrm{j}} \mathrm{p_k} \int_{\mathrm{r_k}}^{\mathrm{r_{k+1}}} r\left(r^2+\Delta^2_{\mathrm{h}}\right)^{-\alpha_{\mathrm{L}}/2}\,\mathrm{d}r \\
&= 2\pi\lambda\mathrm{P_{tx}G_{tot}A_L} \sum_{k=\mathrm{i}}^{\mathrm{j}} \mathrm{p_k} \left[\frac{(r^2+\Delta^2_{\mathrm{h}})^{1-0.5\,\alpha_{\mathrm{L}}}}{2(1-0.5\,\alpha_{\mathrm{L}})}\right]_{r=\mathrm{r_k}}^{\mathrm{r_{k+1}}},
\end{aligned}
$$

yielding the desired result.

The expression for $\sigma^2_{I|R_{\mathrm{S}}}$ can be obtained following a similar rationale.